\documentclass[aip,twocolumn,showpacs,secnumarabic,graphics,floatfix,nofootinbib,tightenlines,nobibnotes,aps,prl,10pt]{revtex4-1}
\usepackage{graphicx}
\usepackage[utf8]{inputenc}
\usepackage[T1]{fontenc}
\usepackage[english]{babel}
\usepackage{array,multirow,makecell}
\usepackage{amsmath}
\usepackage{amssymb}
\usepackage{mathrsfs}
\usepackage{amsmath}
\usepackage[squaren, Gray, cdot]{SIunits}
\usepackage{,amsfonts}
\usepackage{fancyhdr}
\usepackage{colortbl}
\usepackage[nottoc, notlof, notlot]{tocbibind}
\usepackage{vmargin}
\usepackage[colorlinks = true,linkcolor = blue, urlcolor  = blue,citecolor = blue,anchorcolor = blue]{hyperref}
\usepackage{natbib}
\usepackage{float}
\usepackage{url}

\setmarginsrb{2cm}{0.5cm}{2cm}{3cm}{0cm}{1cm}{2cm}{1cm}
\setcellgapes{1pt}
\makegapedcells
\newcolumntype{R}[1]{>{\raggedleft\arraybackslash }b{#1}}
\newcolumntype{L}[1]{>{\raggedright\arraybackslash }b{#1}}
\newcolumntype{C}[1]{>{\centering\arraybackslash }b{#1}}

\begin{document}

\title{\textbf{\Large{The Crystal Structure of Rb$_{2}$Ti$_{2}$O$_{5}$}}}
\author{R. Federicci\textsuperscript{1*}, B. Baptiste$^2$,F.Finocchi$^3$, A. F. Popa $^4$, L. Brohan$^4$, K. Beneut$^2$, P. Giura$^2$, G. Rousse$^{5,6}$, A. Descamps-Mandine$^1$, T. Douillard$^7$,A. Shukla$^2$ and B. Leridon$^1$}
\affiliation{$^1$ \textit{LPEM-ESPCI Paris, PSL Research University, CNRS, Sorbonne Universités UPMC, 10 rue Vauquelin, F-75005 Paris, France}\\ $^2$ \textit{IMPMC, Sorbonne Universités UPMC, CNRS, 4 place Jussieu, F-75005 Paris, France}\\ $^3$ \textit{INSP, Sorbonne Universités UPMC, CNRS, 4 place Jussieu, F-75005 Paris, France}\\ $^4$ \textit{Institut des Matériaux Jean Rouxel (IMN), Université de Nantes, CNRS, 2 rue de la Houssinière, BP 32229, F- 44322 Nantes Cedex 3, France}\\ $^5$ \textit{UMR 8260, "Chimie du Solide et Energie" , Collège de France, Sorbonne Universités  UPMC, 11 Place Marcelin Berthelot, 75231 Paris Cedex 05, France}\\  $^6$ \textit{Réseau sur le Stockage Electrochimique de l'Energie (RS2E), 
FR CNRS 3459, France}\\ $^7$ \textit{Université de Lyon, INSA Lyon, CNRS, MATEIS, UMR 5510, F-69621 Villeurbanne, France}}

\date{\today}

\begin{abstract}

Recent results have demonstrated an exceptionally high dielectric constant in the range 200\,K-330\,K in a crystalline tianium oxide : Rb$_2$Ti$_2$O$_5$. In this article, the possibility of a structural transition giving rise to ferroelectricity is carefully inspected. In particular X-Ray diffraction, high resolution transmission electron microscopy and Raman spectroscopy are performed. The crystal structure is shown to remain invariant and centrosymmetric at all temperatures between 90\,K and 450\,K. The stability of the \textit{C}2/\textit{m} structure is confirmed by DFT calculations. These important findings allow to discard the existence of a conventional ferroelectric phase transition as a possible mechanism for the unusual dielectric constant observed in this material.

\end{abstract}

\maketitle
\section{Introduction}

The investigation of new layered materials has always been topical in condensed matter, since many electronic properties are marginally true in two dimensions. On the other hand, ternary titanium oxydes, such as perovskite, are known to exhibit many remarkable electronic behaviors such as ferroelectricity \cite{Na1.7K0.3Ti3O7, K2Ti6O13-1}, pyroelectricity \cite{Pyro-RTPO-2006, Pyro-KTAO-2010}, colossal magnetoresistance \cite{CMR} or 2D-superconducting properties \cite{LAOSTO}. The ternary titanium oxide family denoted by M$_{2}$Ti$_{n}$O$_{2n+1}$ (MTO) with M = Li, Na, K, Rb, Cs, Fr and called the Anderson-Wadsley type alkali titanates \cite{K2Ti2O5-1986} with layered structure for 2 $\le n \le$ 5 and tunnel structure for 6 $\le n \le$ 8, are known for presenting ferroelectric phase transitions, with especially high Curie temperatures\cite{K2Ti6O13-1} as well as for other applications \cite{Shripal2005}. Their ease of synthesis and their potential applications are a motivation to study this family of compounds. \newline
Very few information has been reported to date in the literature concerning the Rb$_{2}$Ti$_{2}$O$_{5}$ (RTO) compound, which is reputed to be a layered material. Actually, two early documents report its synthesis and claim a space group \textit{Cm} for its structure \cite{Rb2Ti2O5-1, Rb2Ti2O5-2}. \newline
Earlier structural investigations were led on K$_{2}$Ti$_{2}$O$_{5}$, where the space group is reported to be \textit{C}2/\textit{m} and more generally on the M$_{2}$Ti$_{2}$O$_{5}$ material family (MTO)\cite{K2Ti2O5-1960, K2Ti2O5-1961,K2Ti2O5-1986}.  
The material exhibits a five-fold coordination for titanium, which comes as a surprise since titanium atoms tend to adopt an energetically more stable six-fold coordination in most materials \cite{K2Ti2O5-1960}. As a matter of fact, the five-fold Ti coordination has only been reported in MTO. According to the literature, this family of compounds offers a certain interest for chemists because of their oxidation abilities \cite{K2Ti2O5-2009, K2Ti2O5-2011} but electrical properties measurements have not been reported so far. \newline
Recent work \cite{1-2016-RTO-Ionic} has demonstrated superionic conductivity properties attesting of an electrolyte nature of the material due to ionic electromigration. The ionic conductivity in the temperature range between 200\,K and 330\,K reaches as high as $10^{-3}\,S.cm^{-1}$. This high ionic conductivity coupled with an electronically insulating nature leads to an electrolyte behavior. 
These two properties were demonstrated to yield important polarization phenomena with colossal dielectric constant reaching $10^{9}$ at low frequency \cite{1-2016-RTO-Ionic}. \newline
However since these observations are also consistent with ferroelectric like behavior, the question of a possible associated structural phase transition arises, as an alternative explanation. \newline
In the present work, we report a systematic characterization of the Rb$_2$Ti$_2$O$_5$ crystal structure as a function of temperature. X-ray diffraction and Raman spectroscopy techniques were both used in conjunction with theoretical calculations of structural and vibrational properties, using density functional theory.

\section{Chemical synthesis}

The synthesis of the Rb$_{2}$Ti$_{2}$O$_{5}$ was made under air by melting TiO$_{2}$ (rutile) and RbNO$_{3}$ powders in a crucible. 

It resulted hard aggregates of transparent needles that were kept under dry argon atmosphere in order to avoid hydratation. \newline
The crystals extracted from the aggregate exhibit a needle-shape with typically 1 millimeter length and hundreds of micrometers width and thickness. \newline

\section{Characterization}

\subsection{Single crystal XRD}

Single-crystal X-ray diffraction data were acquired on an Oxford Diffraction Xcalibur-S diffractometer equipped with a Sapphire CCD-detector with Mo K$\alpha_1$ radiation ($\lambda$ = 0.71073\,$\angstrom$, graphite monochromator) at 293\,K. Data reduction, cell refinement, space group determination, scaling, and empirical or analytical absorption correction were performed using CrysAlisPro software \cite{DRX1}. \newline
Aquisitions were realized at 400\,K, 300\,K and 150\,K and they all show the same structure. The atomic positions and the lattice parameters of the Rb$_{2}$Ti$_{2}$O$_{5}$ structure are respectively reported in the table \ref{DRX} and \ref{DRXAtomic}. \newline

\begin{table*}[t]
\begin{tabular}{||c|c|c|c||}
\hline\textbf{Temperature} & \textbf{400\,K} & \textbf{300\,K} & \textbf{150\,K} \\
\hline CCDC number &   1517161	& 1517160	& 1517159  \\
\hline Molecular Weight (g.mol$^{-1}$) & 346.74 	& 346.74 	& 346.74  \\
\hline Cryst syst & monoclinic			& monoclinic		& monoclinic \\
\hline Space group & \textit{C}2/\textit{m}	& \textit{C}2/\textit{m} & \textit{C}2/\textit{m} 	\\
\hline a/$\angstrom$ & 11.3419(12)	&	11.3370(13)	&	11.3457(12)	\\
\hline b/$\angstrom$ & 3.8198(5) 	&	3.8244(5)		&	3.8195(5)	 \\
\hline c/$\angstrom$ & 7.0103(8) 	&	6.9946(8)		&	6.9699(8)	\\
\hline $\alpha$/deg	& 90				&	90				&	90	 \\
\hline $\beta$/deg 	& 100.298(11)	&	100.308(12)	&	100.359(11)	\\
\hline $\gamma$/deg & 90			&	90				&	90 \\
\hline Volume/$\angstrom^{3}$ & 298.82(6) 	&	298.37(6)	&	297.12(6)\\
\hline Z & 2 	& 2	& 2\\
\hline $\rho_{calc}$ (mg.mm$^{-3}$) & 3.854 	& 3.860 	& 3.876 \\
\hline $\mu$ (mm$^{-3}$) & 18.796  	& 18.824	& 18.903\\
\hline F(000)  & 316.0  	& 316.0	& 316.0\\
\hline Crystal size & 0.2 $\times$ 0.1 $\times$ 0.05 	& 0.2 $\times$ 0.1 $\times$ 0.05	& 0.2 $\times$ 0.1 $\times$ 0.05\\
\hline Radiation  & MoK$\alpha$ ($\lambda$ = 0.71073) 	& MoK$\alpha$ ($\lambda$ = 0.71073)	& MoK$\alpha$ ($\lambda$ = 0.71073)\\
\hline 2$\Theta$ range   & 5.906 to 64.994 $\degree$ 	& 5.92 to 52.726 $\degree$ 	& 5.942 to 52.714 $\degree$\\
\hline Reflexions collected  & 1859 	& 1389	& 1205\\
\hline Data/restraints/param.   & 567/0/30& 351/0/29	& 348/0/29\\
\hline GOF on F$^2$    & 1.081	& 1.173	& 1.109\\
\hline R1/wR2 (I>=2$\sigma$ (I))   & 0.0353 /0.0893	& 0.0279 /0.0714	& 0.0499/0.1310\\
\hline R1/wR2 (all)  & 0.0400/0.0918		& 0.0293/0.0725	& 0.0537/0.1342\\
\hline Largest diff. peak/hole / e$\angstrom^{3}$   & 0.82/-1.63 	& 0.68/-1.45	&1,52/-1.93\\
\hline
\end{tabular}
\caption{Measurements Conditions and Crystallographic Data for Room-Temperature Single-Crystal X-ray Diffraction on Rb$_{2}$Ti$_{2}$O$_{5}$ under nitrogen at 400\,K, 300\,K  and150\,K.}
\label{DRX}
\end{table*}

\begin{table}[h!]
\centering
\begin{tabular}{||c|c|c|c|c|c||}
\hline 		&\textbf{Atom} & \textbf{x} & \textbf{y} & \textbf{z} & \textbf{U(eq)}\\
\hline
\hline \multirow{6}{*}{400\,K} &  Rb1 & 981.2(8) & 5000  &   8472.5(12)	 & 14.9(5)\\\cline{2-6}
		& Ti1 & 3522.9(14) & 5000  &   5897(2) & 9.7(5) \\\cline{2-6}
	& O1 & 5000 & 5000  &   5000 & 15(2)\\\cline{2-6}
		& O2 & 3767(6) & 5000  &   8391(9) & 15.6(15)\\\cline{2-6}
		& O3 & 1772(6) & 5000  &   4776(9) & 14.9(14)  \\\cline{2-6}
\hline
\hline \multirow{6}{*}{300\,K} &  Rb1 & 985.1(5) & 5000  &   8481.2(8) & 21.0(3)\\\cline{2-6}
		& Ti1 & 3524.9(8) & 5000  &   5884.9(14) & 10.8(3) \\\cline{2-6}
	& O1 & 5000 & 5000  &   5000 & 19.7(13)\\\cline{2-6}
		& O2 & 3762(4) & 5000  &   8367(6) & 20.7(9)\\\cline{2-6}
		& O3 & 1780(3) & 5000  &   4797(5) & 14.3(8)  \\\cline{2-6}
\hline
\hline \multirow{6}{*}{150\,K} &  Rb1 & 981.2(8) & 5000  &   8472.5(12)	 & 14.9(5)\\\cline{2-6}
		& Ti1 & 3522.9(14) & 5000  &   5897(2) & 9.7(5) \\\cline{2-6}
		& O1 & 5000 & 5000  &   5000 & 15(2)\\\cline{2-6}
		& O2 & 3767(6) & 5000  &   8391(9) & 15.6(15)\\\cline{2-6}
		& O3 & 1772(6) & 5000  &   4776(9) & 14.9(14)   \\\cline{2-6}
\hline
\end{tabular}
\caption{Fractional Atomic Coordinates ($\times$10$^4$) and Equivalent Isotropic Displacement Parameters (\AA $^2 \times$10$^3$) for Rb$_2$Ti$_2$O$_5$. Ueq is defined as 1/3 of of the trace of the orthogonalised U$_{IJ}$ tensor.}
\label{DRXAtomic}
\end{table}


\subsection{Powder XRD}

\begin{figure}[h!]
\begin{center}
\includegraphics[width=8.2cm, trim = 2cm 1.2cm 3cm 2cm, clip]{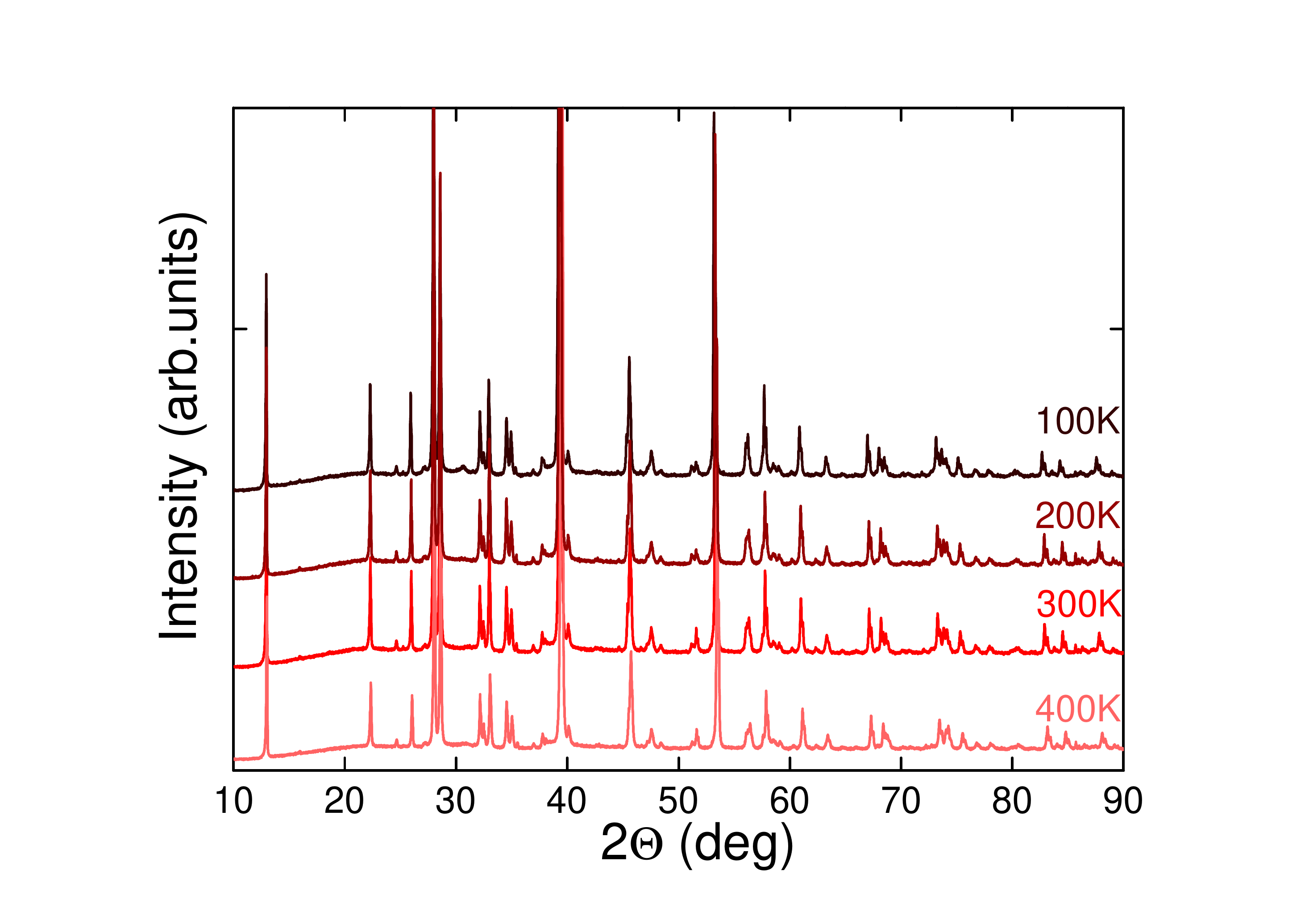}
\caption{Diffractograms acquired on Rb$_{2}$Ti$_{2}$O$_{5}$ powder as function of temperature between 100\,K and 400\,K.}
\label{DRX-Poudre}
\end{center}
\end{figure}

The X-ray powder diffraction (XRD) patterns were recorded at the IMPMC (Pierre and Marie Curie University, Paris) x-ray diffraction platform using an XPert Pro Panalytical diffractometer equipped with a Cu K$\alpha$ radiation source ($\lambda_{K \alpha 1}$ = 1.54056\,$\angstrom$, $\lambda_{K \alpha 2}$ = 1.54439\,$\angstrom$ ) with an XCelerator detector. The measurements were conducted under nitrogen atmosphere (under vacuum for 100\,K acquisition) in an Anton Paar HTK 450 temperature controlled chamber. Rietveld refinements \cite{DRX4} were performed with the FullProf suite of programs \cite{DRX5} on X-ray powder diffraction patterns recorded in Bragg-Brentano geometry. The background was modelized using a 12-coeficients polynom, and Thomson-Cox-Hasting Voigt functions \cite{DRX6} were used as peak shape functions. Preferred orientation, consistent with a platy habit of crystallites, was taken into account in the refinements as all (001) reflections were over-estimated. Measurements in capillary would have limited this effect but they were not possible since the Bragg Brentano (reflection) geometry was imposed by our cryostat. However, these refinements reveal that the single crystal structure is representative of the whole sample and further confirm the absence of structural transition from 100\,K to 400\,K. Vesta software \cite{VESTA} was used to visualize the crystal structures. These measurements were made as function of temperature and under a nitrogen atmosphere. \newline
The evolution of the diffractograms acquired on a RTO is pictured in figure \ref{DRX-Poudre} as function of temperature between 100\,K and 400\,K. Table \ref{PowderData} displays the lattice parameters deduced from Rietveld refinements performed on the X-ray Powder Pattern. \newline

\begin{table*}[t]
\begin{tabular}{||c|c|c|c|c||}
\hline  \textbf{Temperature}& \textbf{400\,K} & \textbf{300\,K} & \textbf{200\,K} & \textbf{100\,K}\\
\hline Space group & \textit{C}2/\textit{m} & \textit{C}2/\textit{m} &  \textit{C}2/\textit{m} &  \textit{C}2/\textit{m}\\
\hline a\,($\angstrom$) & 11.34378(2) & 11.34378(2) &  11.34381(2) &  11.34265(3)\\
\hline b\,($\angstrom$) & 3.82824(5) & 3.82905(6) &  3.82756(5) & 3.82557(6)\\
\hline c\,($\angstrom$) &  6.99109(5) & 7.00371(5) &  6.98718(5)  &  6.96969(6)\\
\hline $\beta$\,(deg) & 100.2889(5) & 100.2413(6) & 100.2939(6) & 100.3379(7)\\
\hline Volume\,($\angstrom^3$) & 298.762(7) & 299.365(8) &  298.495(7)& 297.521(9)\\
\hline R$_{Bragg}$\,(\%)  & 9.86 & 13.4 &  12.1 & 13\\
\hline
\end{tabular}
\caption{Results of the Rietveld Refinement on the X-ray Powder Pattern of Rb$_{2}$Ti$_{2}$O$_{5}$ at different temperatures.}
\label{PowderData}
\end{table*}

\begin{figure}[h!]
\begin{center}
\includegraphics[width=10cm, trim = 4cm 6cm 3cm 5cm, clip]{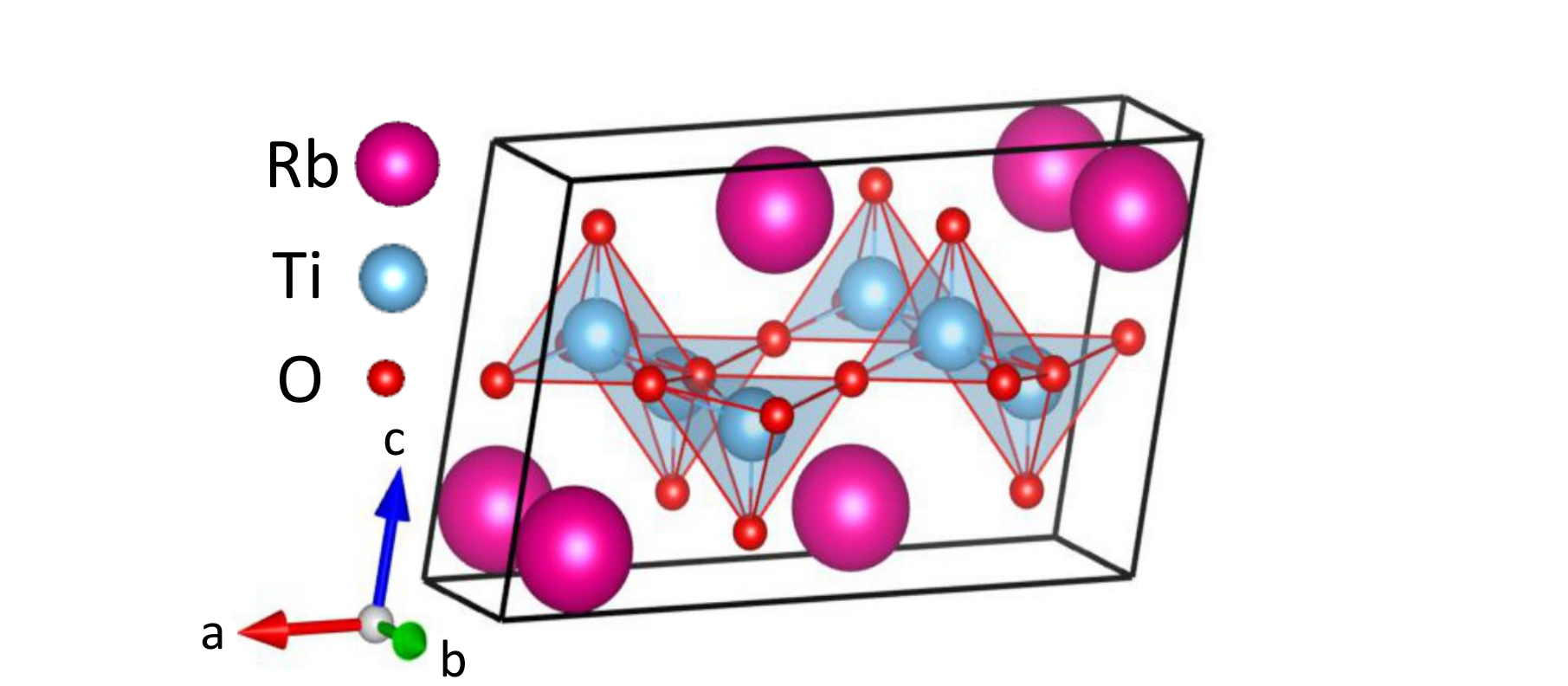}
\caption{Structure of Rb$_{2}$Ti$_{2}$O$_{5}$ visualized by VESTA \cite{VESTA} software. The Rb atoms are shown in pink, the Ti atoms in blue and the O atoms in red.}
\label{RTO-cell}
\end{center}
\end{figure}

The atomic positions and lattice parameters extracted from the single crystal and the powder measurements were found to be similar and allow to conclude with a \textit{C}2/\textit{m} structure which remains unchanged over the investigated range of temperature. \newline
The unit cell content of the RTO is shown in figure \ref{RTO-cell} using VESTA software. \newline

The structure of RTO is highly anisotropic. This two-dimensional structure consists in alternate layers of Ti-O and Rb atoms along the c-axis. The Ti-O layers are made of chains of double pyramids sharing corners. A double-pyramid is formed by the association of two square-based pyramids sharing edges, with their apex pointing to opposite directions. In this configuration each Ti atom is surrounded by five oxygens which form a slightlty tilted square based pyramid. Rb$^+$ atoms sit in between those layers. The refinements performed demonstrated that the structure is identical with the one reported for K$_{2}$Ti$_{2}$O$_{5}$ \cite{K2Ti2O5-1961}.\newline

\subsection{Scanning electron microscopy}

\begin{figure}[h!]
\begin{center}
\includegraphics[width=10cm, trim = 6cm 1cm 4cm 0.5cm, clip]{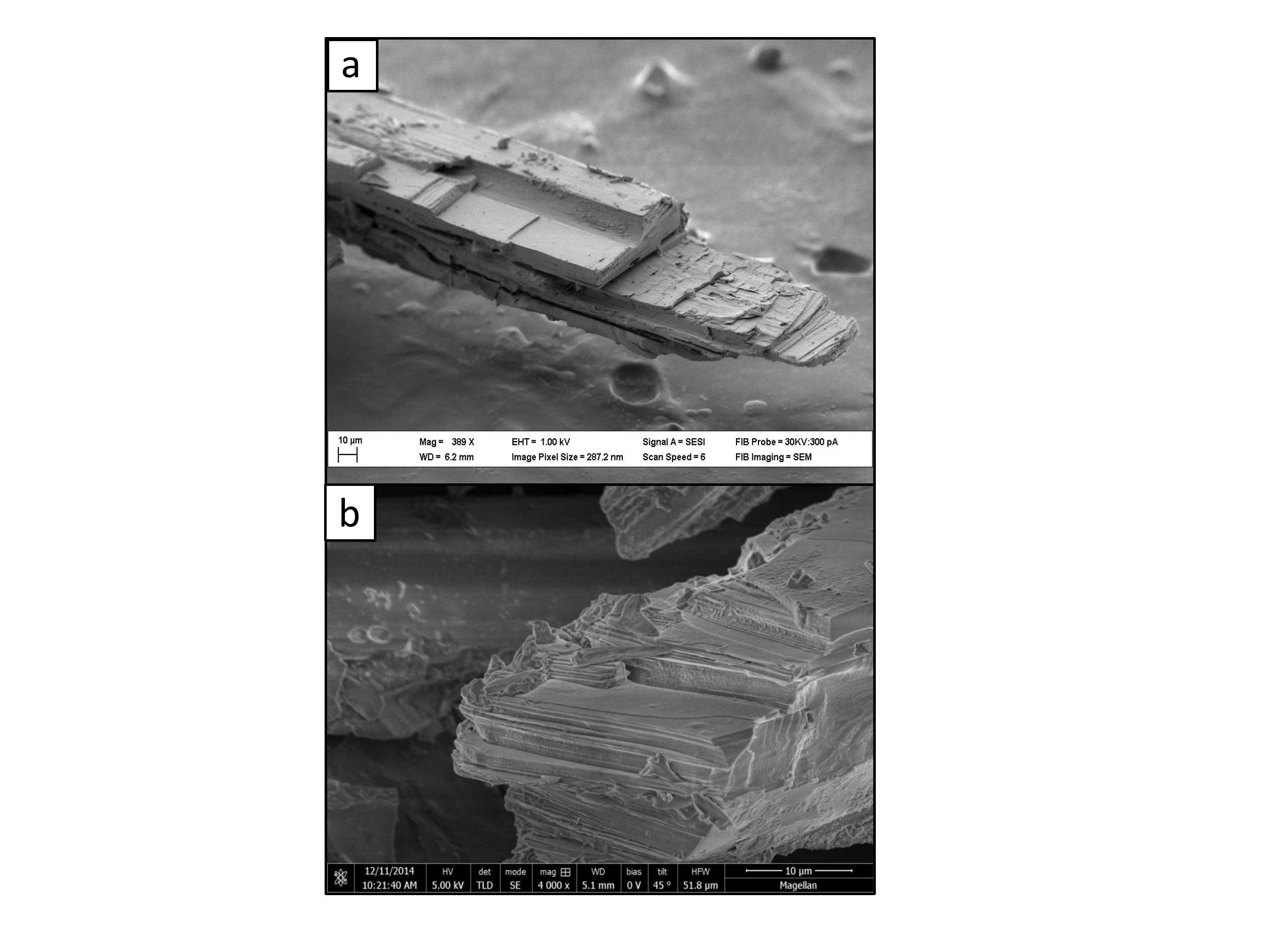}
\caption{SEM pictures acquired on two different Rb$_{2}$Ti$_{2}$O$_{5}$ crystals from different batches isolated just after the chemical synthesis  process reached back room temperature.}
\label{RTO-MEB}
\end{center}
\end{figure}

Rightafter the chemical synthesis process reached back room temperature, RTO cristals were isolated and Scanning Electron Microscopy (SEM) acquisitions were realized. Two SEM pictures which show the needle shape of the RTO and its lamellar structure are displayed in figure \ref{RTO-MEB}. \newline
The growth direction is along the b axis. In these pictures the microscopic layered nature of the material is observable at the macroscopic scale over the entire size of the cristal. Figure \ref{RTO-MEB}.a displays a good overview of the superimposition of layers. \newline

\subsection{High-resolution transmission electron microscopy}

\begin{figure}[h!]
\begin{center}
\includegraphics[width=1.0\linewidth, trim = 0cm 0cm 0cm 0cm, clip]{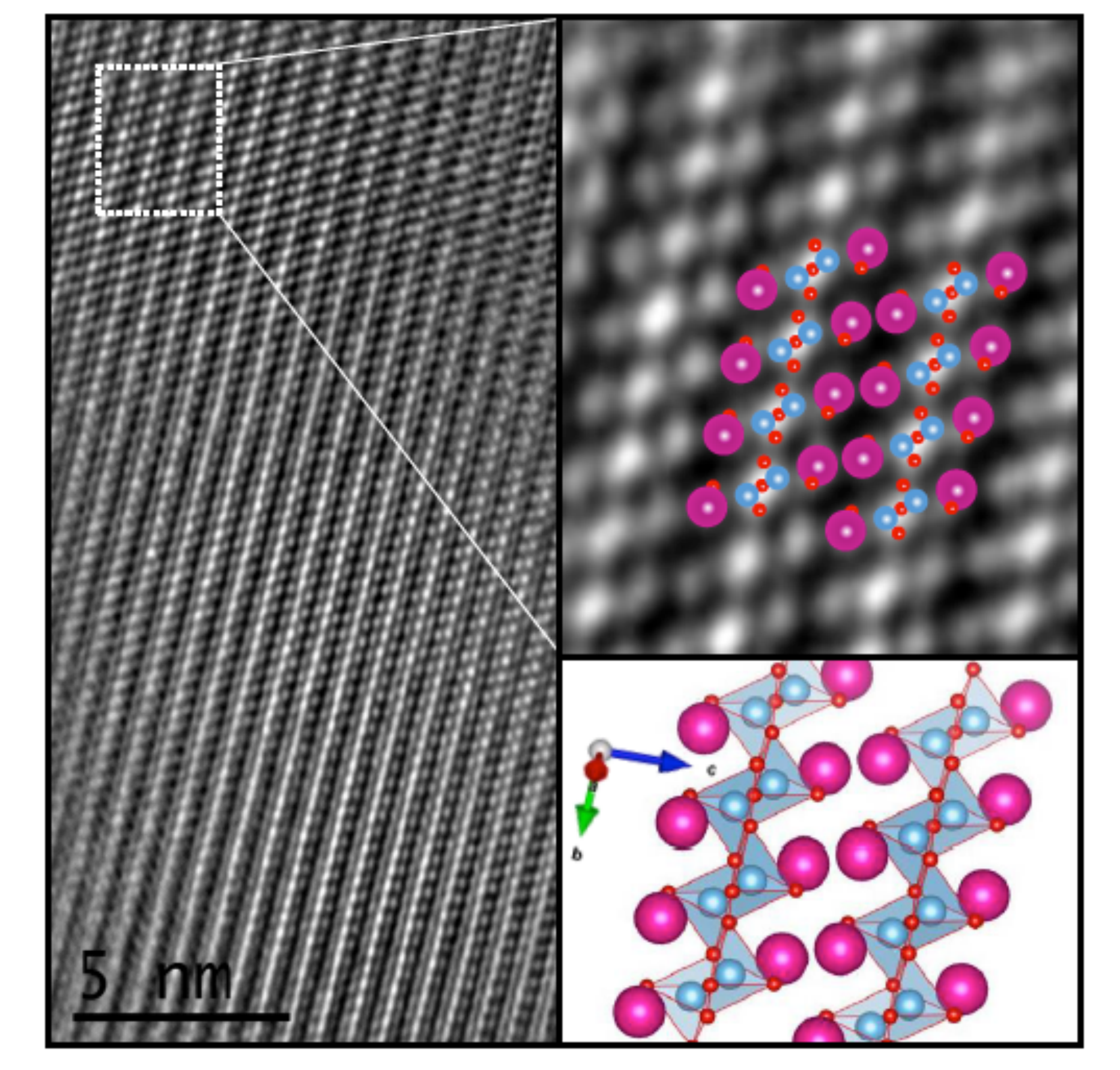}
\end{center}
\caption{High resolution transmission electron microscopy image acquired on a Rb$_{2}$Ti$_{2}$O$_{5}$ sample along the [110] zone axis (HR-TEM JEM-2010F; JEOL Ltd,  Tokyo, Japan). A close-up of the image is showed in the insert on which is  superimposed a simulated crystal structure from X-Ray diffraction data. The Rb atoms are in pink, the Ti atoms are in blue and the O atoms are in red.}
\label{RTO-TEM}
\end{figure}

Pictures of High-resolution transmission electron microscopy (HRTEM) were performed on a RTO sample (HR-TEM JEM-2010F; JEOL Ltd,  Tokyo, Japan). The figure \ref{RTO-TEM} pictured one of those acquisitions. \newline

A close-up of the crystal ordering is shown in the inset of figure \ref{RTO-TEM}. A simulated crystal structure realized by VESTA is surimposed over the HRTEM image pointing to the crystal of the studied sample. \newline
The zig-zag light spot lines represent the (ab) plane made of the Ti-O squared base-pyramids as can attest the insert surimposed structure. Very shady lights in between these light lines show the presence of rubidium atoms forming the layers separating the Ti-O plans. \newline

\section{Raman spectroscopy}

\begin{figure*}[t]
\center
\includegraphics[width=17cm, trim = 2cm 10.5cm 2cm 0cm, clip]{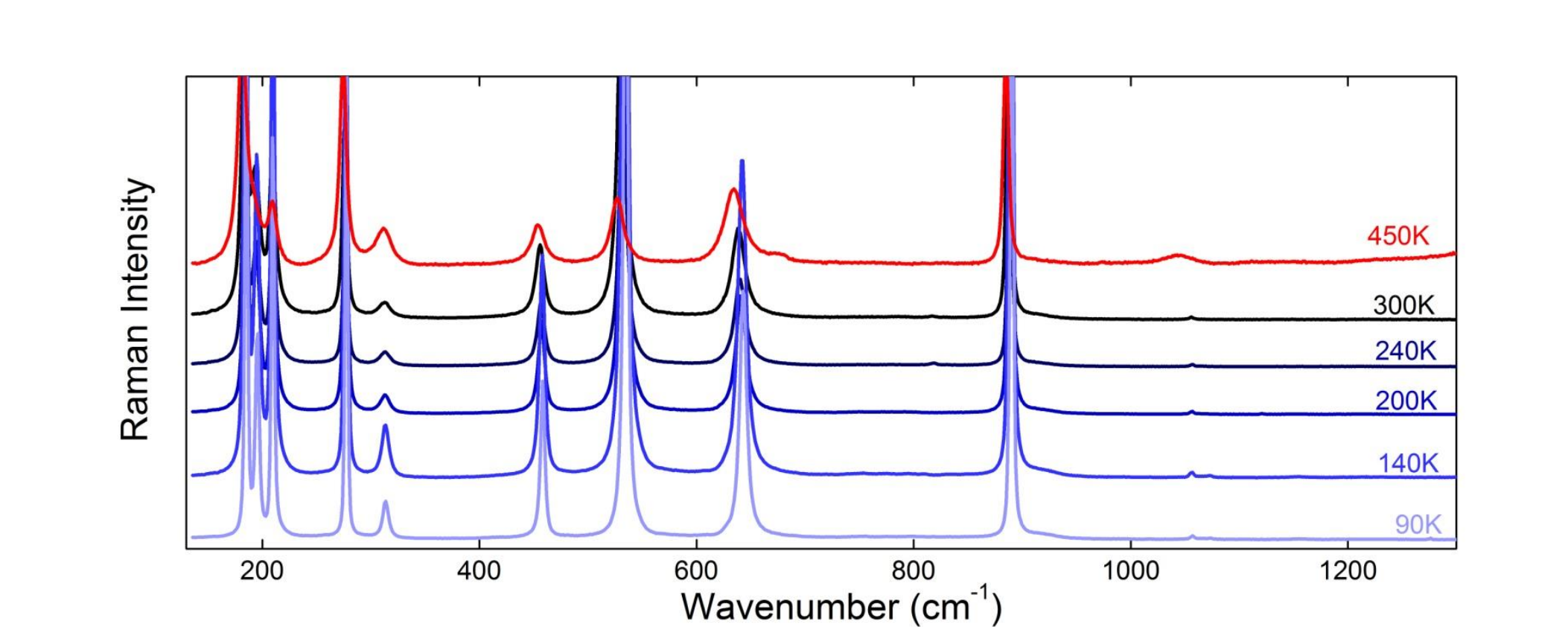}
\caption{Experimental Raman spectra acquiered on a Rb$_2$Ti$_2$O$_5$ crystal as function of the temperature under Helium atmosphere between 180\,cm$^{-1}$ and 1300\,cm$^{-1}$. No significative changes happen when temperature decreases.}
\label{RTO-Raman-Temp}
\end{figure*}

Raman spectroscopy measurements were performed on Rb$_2$Ti$_2$O$_5$ cristals over the range of temperature [90\,K - 450\,K] under controlled atmosphere conditions. The set-up consisted of an argon laser source "Spectra-physics Argon 514.5\,nm", a spectrometer "T64000 HORIBA Jobin Yvon", a set of optical microscopy lens used to focus the laser beam over few micrometers and a cryostat or a cryofour "linkam FTIR 600" used to control the atmosphere and the temperature with helium. The power of the laser beam was set at 140\,mW and the recording times lasted 30\,sec. Each Raman spectra is an average from three distinct spectra acquired in a row. \newline

Raman spectra recorded between 90\,K and 450\,K over the wavenumber range [180\,cm$^{-1}$ - 1300\,cm$^{-1}$] are plotted in figure \ref{RTO-Raman-Temp}. The spectra remain qualitatively unchanged when temperature varies attesting of the stability of the structure over the entire explored range of temperature. Measurements were also performed at 90\,K and did not show any variation of the phonon modes. This Raman spectroscopy investigation represents therefore a second probe attesting of the stability of the \textit{C}2/\textit{m} crystal structure of the RTO material between 10\,K and 450\,K.

\section{Density functional theory calculation}

\subsection{Crystal structure}

The structural calculations were performed on the Rb$_2$Ti$_2$O$_5$ unit cell, 
which consists of two formula units, through the Quantum Espresso package \cite{QE} that is based on the density functional theory (DFT). Both Local Density Approximation (LDA) and Generalized Gradient Approximation (GGA) functionals were used. \cite{PBE} 
\newline

\begin{table}[h!]
\begin{tabular}{||c|c|c||}
\hline	&GGA & LDA\\
\hline	a/$\angstrom$ & $11.3318 $  & $ 11.1425$ \\
\hline	b/$\angstrom$ & $3.8322$ & $3.7492$ \\
\hline	c/$\angstrom$ &$6.9863 $ &  $ 6.7665$ \\
\hline	$\alpha$/deg & 90  & 90 \\
\hline	$\beta$/deg & 100.110  & 100.327\\
\hline	$\gamma$/deg & 90 &  90\\
\hline	$V_0$/$\angstrom^{3}$  & 298.688  & 278.105 \\
\hline	$E_0$/Ryd & -980.1266(6) & -980.4192(1)\\
\hline	$B_0$/GPa & 53.4(0)  & 70.9(8)  \\
\hline	$B_0'$  & 7.0(5) & 4.5(9) \\
\hline
\end{tabular}
\caption{Computed equilibrium lattice constants a, b, c and angles $\alpha$, $\beta$, $\gamma$ 
within the GGA or the LDA.  
The equilibrium volume $V_0$, bulk modulus $B_0$ and its pressure derivative $B_0'$
have been obtained through a fit to the Murnaghan equation of state (see text). }
\label{DFT}
\end{table}

We employed ultra-soft pseudopotentials, which include semi-core states for Rb and Ti. A Monkhorst-Pack $2 \times 6 \times 3$ grid and a cutoff energy of 50\,Ryd for the expansion of the Kohn-Sham \cite{KohmSham1, KohmSham2} orbitals in plane waves (4 times as large for the charge density and potential) were enough to obtain the converged energies and the structural parameters. \newline

At the beginning of the structural optimization, the space group was set as \textit{C}2/\textit{m} and the atomic positions were those extracted from the X-Ray diffraction measurements, which are displayed in Table \ref{DRXAtomic}. Both lattice parameters, angles and atomic positions were then left free to relax; their final values, within the GGA or the LDA, are collected in Table \ref{DFT}. A series of structural optimizations have been conducted at several volumes around the equilibrium volume and the numerical results have been represented through the Murnaghan equation of state\cite{Murnaghan3}, via the fitted equilibrium volume $V_0$, bulk modulus  $B_0$ and its pressure derivative $B_0'$. \newline

\begin{table}[h!]
\begin{tabular}{||c|c|c|c||}
\hline	 \textbf{Atom} & \textbf{x} & \textbf{y} & \textbf{z}\\
\hline	 {Rb} &1003  & 5000     & 8500   \\
\hline	 {Ti} & 3521  &5000    & 5900  \\  
\hline	 {O1} & 5000 & 5000  & 5000 \\
\hline	 {O2} & 3745    &5000    &  8400    \\
\hline	 {O3} &  1776      & 5000     & 4787    \\
\hline
\end{tabular}
\caption{Fractional Atomic Coordinates ($\times$10$^4$) after structural relaxation for the Rb$_2$Ti$_2$O$_5$ crystal, within the GGA.}
\label{GGAAtomic}
\end{table}


The computed atomic positions are reported in table \ref{GGAAtomic}. We see that the structural parameters that were computed within the GGA are extremely close to their experimental counterparts that are collected in Tab \ref{DRX}. The computed enthalpy of formation of Rb$_2$Ti$_2$O$_5$ at zero pressure from the elementary stable phases Rb$_{\rm (s)}$, Ti$_{\rm (s)}$ and O$_{2{\rm (gas)}}$ is -309.8 kJ/mol, within the GGA. These theoretical calculations thus confirm that the synthesized Rb$_2$Ti$_2$O$_5$ crystal is thermodynamically stable and that its structure can be solved with a high level of accuracy by combining X-ray diffraction with DFT calculations.  
\newline 

\subsection{Band structure}

\begin{figure}[h!]
\begin{center}
\includegraphics[width=8.3cm, trim = 2cm 2cm 2cm 2cm, clip]{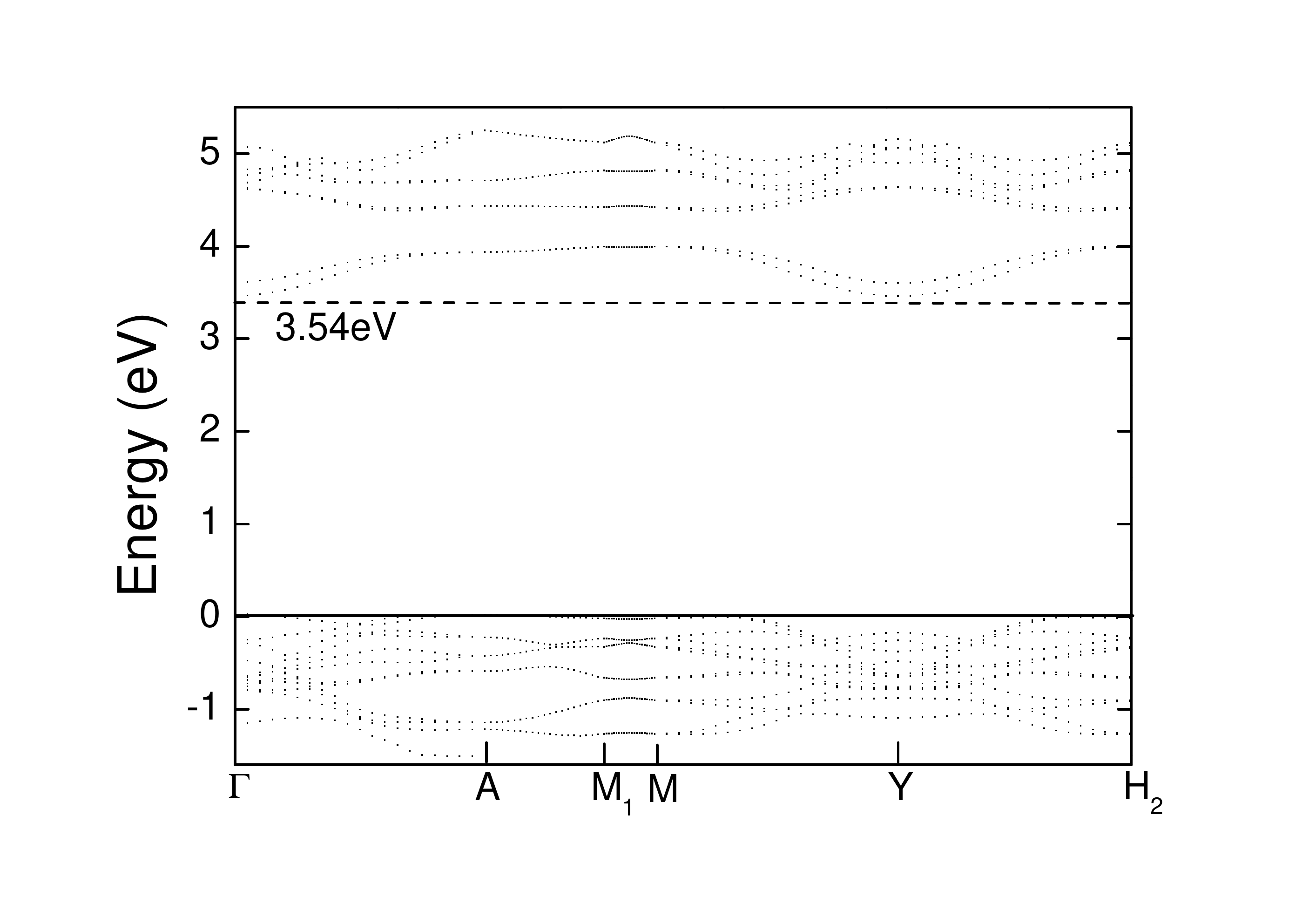}
\caption{Band structure of Rb$_2$Ti$_2$O$_5$ calculated with GGA functional. The high-symmetry points are chosen following the path of the MCL structure\cite{DFTPath}. The highest occupied electronic level in the valence band is set at zero, while the dashed line makes the lowest electronic level in the conduction band.}
\label{RTO-band}
\end{center}
\end{figure}

The band structure pictured in the figure \ref{RTO-band} is calculated using the GGA functional for the corresponding optimized crystal structure. The path used along the Brillouin zone is chosen using the MCL structure \cite{DFTPath}. A gap of at least 3.54\,eV, which is likely underestimated as usual in the DFT-GGA, attests nevertheless of the insulating nature of the Rb$_2$Ti$_2$O$_5$ crystal. \newline

\subsection{Phonon modes}

\begin{figure*}[t]
\includegraphics[width=17cm, trim = 4.5cm 0.8cm 4cm 0cm, clip]{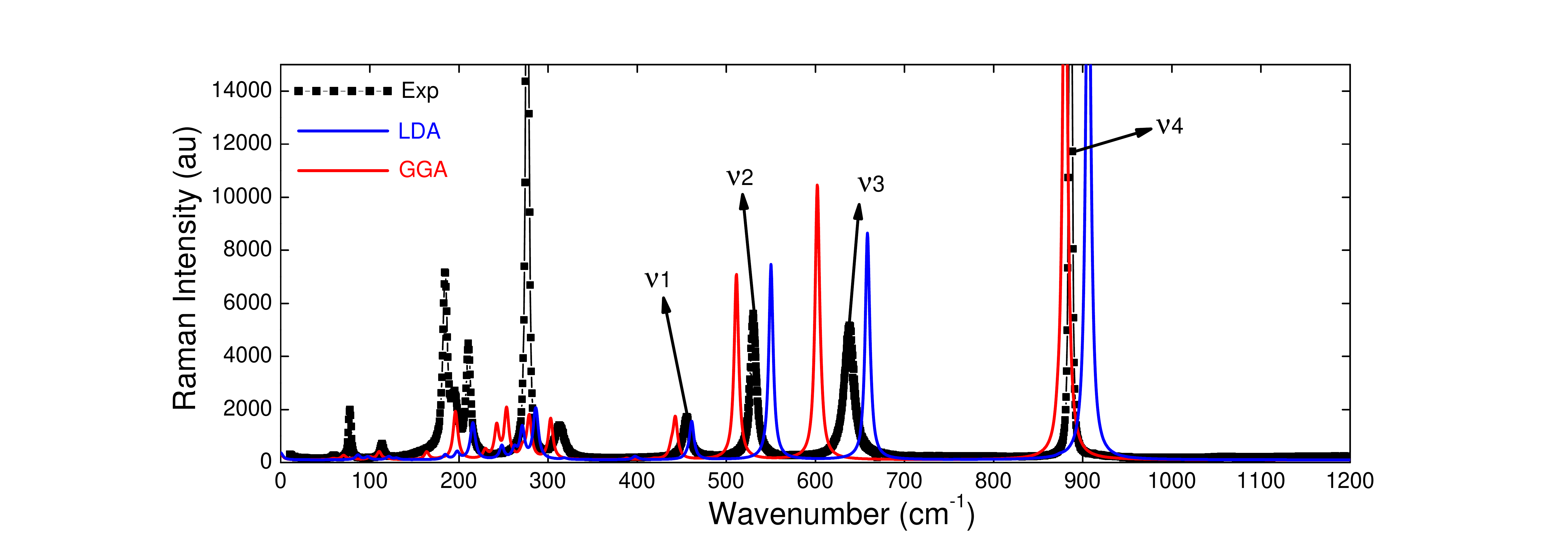}
\caption{Experimental and simulated Raman spectra. The black square curve is the experimental Raman spectrum between 10\,cm$^{-1}$ and 1200\,cm$^{-1}$ acquiered on a crystal of Rb$_{2}$Ti$_{2}$O$_{5}$ under room conditions. The blue curve is the theoretical raman spectrum simulated with DFT using LDA functional and the red curve using GGA functional. }
\label{RTO-Raman-300K}
\end{figure*}

\begin{figure*}[t]
\includegraphics[width=16cm, trim = 4cm 0.5cm 3cm 0.1cm, clip]{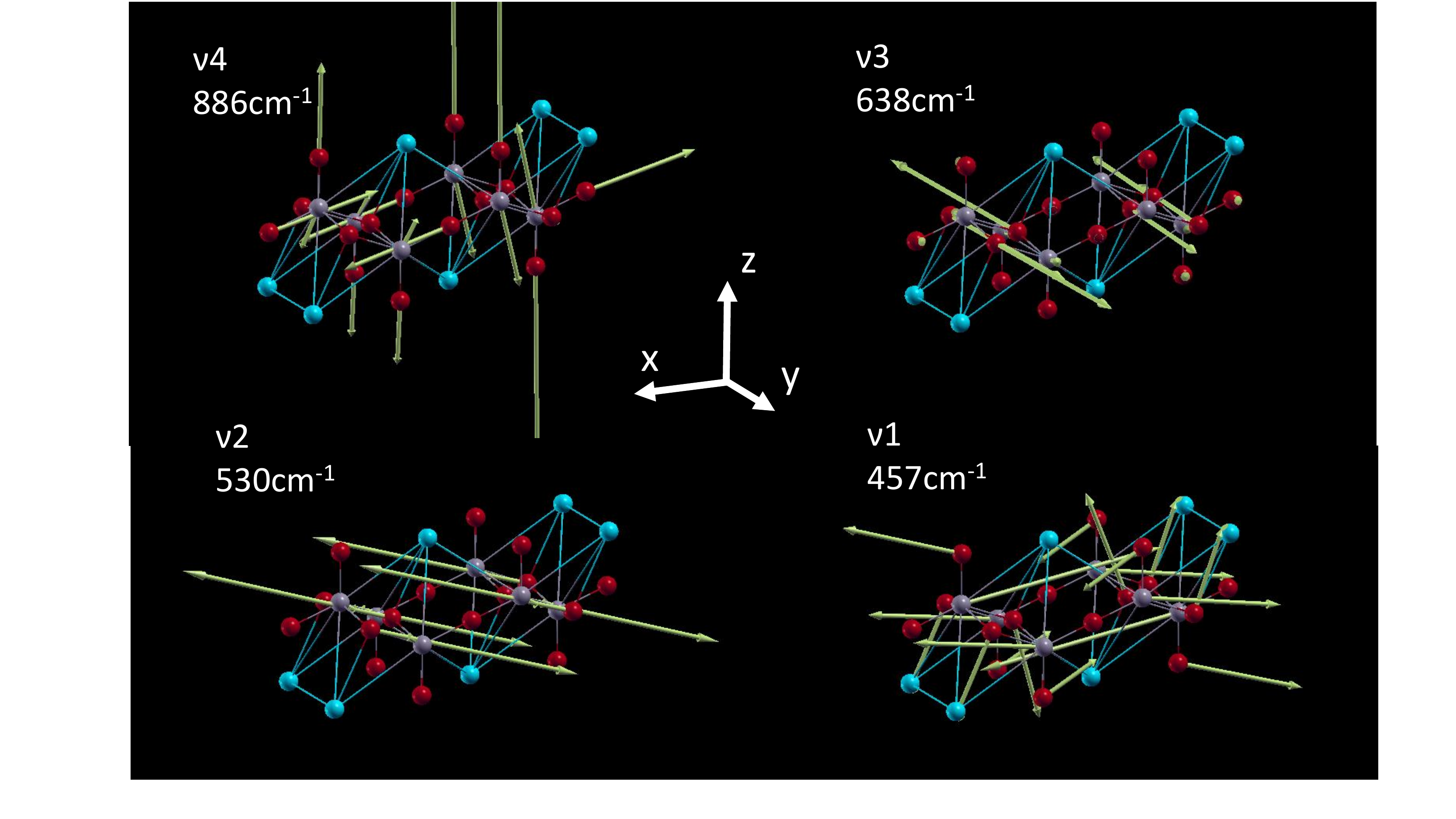}
\caption{Representations of the vibrational phonon modes N$\degree$ 54, 46, 45 and 42 calculated with density function theory for Rb$_{2}$Ti$_{2}$O$_{5}$ system using generalized gradient approximation. The Raman intensities are displayed in Appendix B in the table \ref{DFT-PBE-PhononFrequency}}
\label{PhononModes}
\end{figure*}

\begin{figure*}[t]
\includegraphics[width=16cm, trim = 0.5cm 1cm 1.6cm 0.2cm, clip]{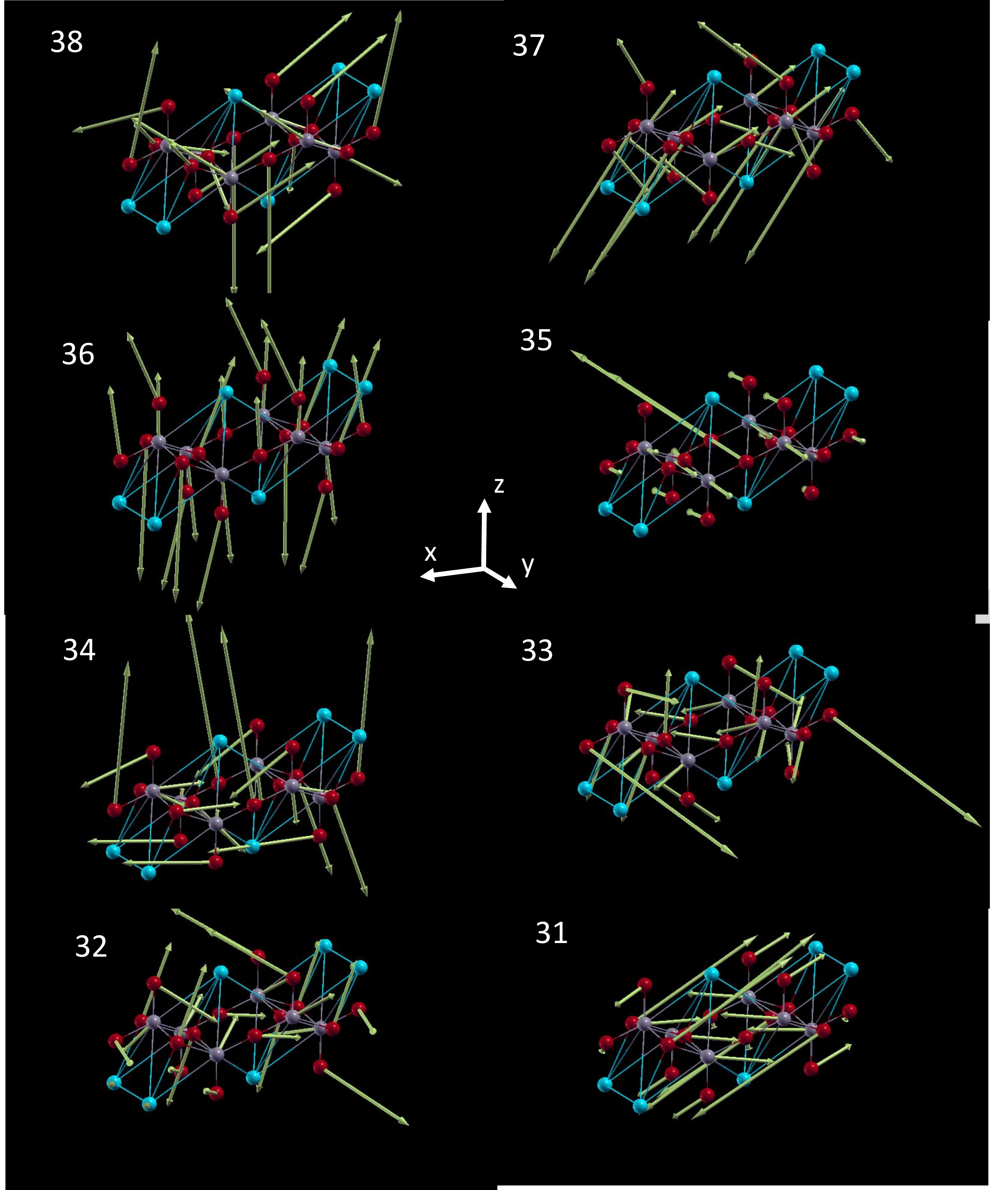}
\caption{Representations of the vibrational phonon modes N$\degree$ 38, 37, 36, 35, 34, 33, 32 and 31 calculated using Density Function Theory for Rb$_{2}$Ti$_{2}$O$_{5}$ system. The frequency and Raman intensity of the related phonon modes are displayed in the table \ref{DFT-PBE-PhononFrequency}}
\label{Phonon31-38}
\end{figure*}
\begin{figure*}[t]
\includegraphics[width=16cm, trim = 0.5cm 0.5cm 0.8cm 0.5cm, clip]{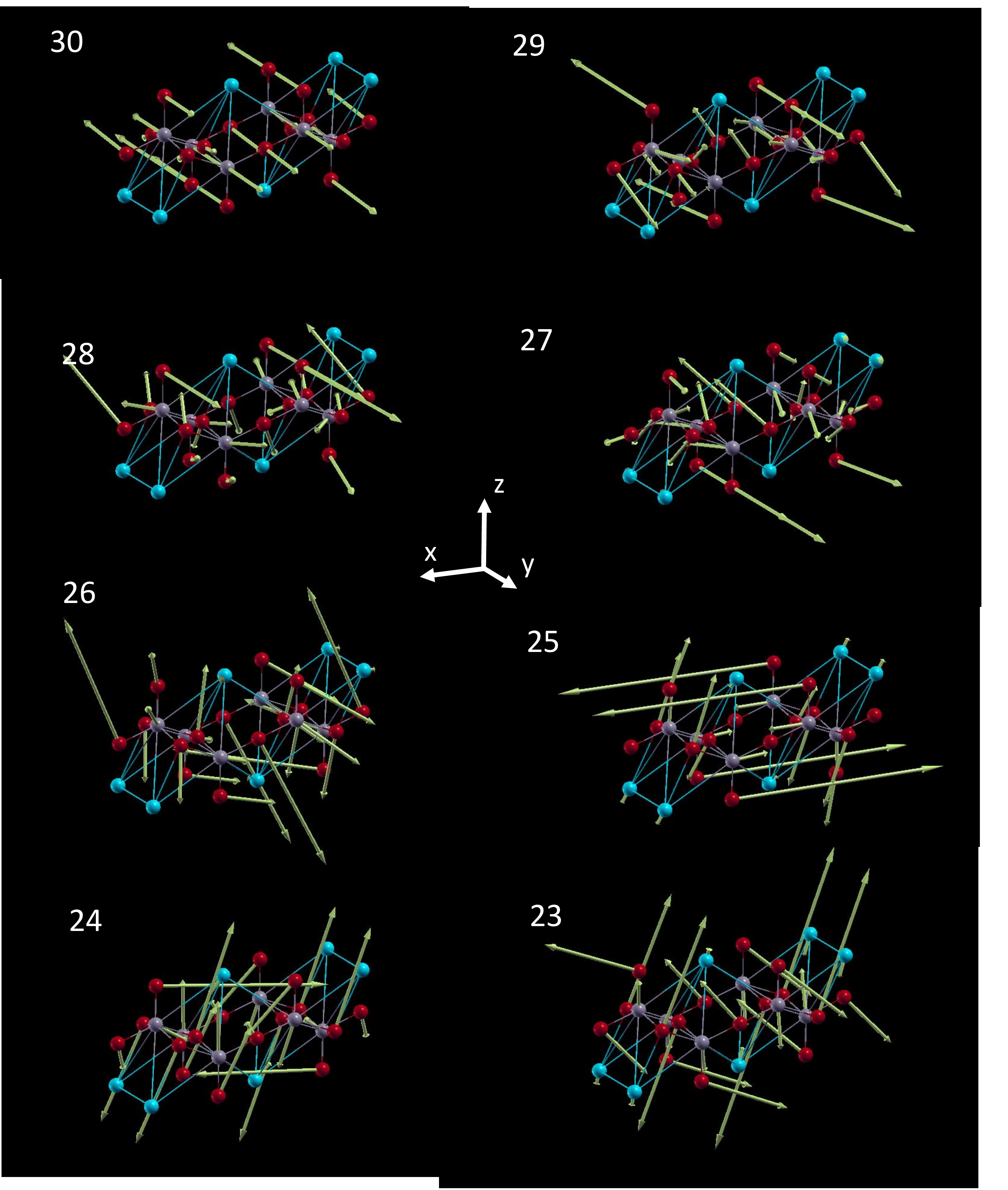}
\caption{Representations of the vibrational phonon modes N$\degree$ 30, 29, 28, 27, 26, 25, 24 and 23 calculated using Density Function Theory for Rb$_{2}$Ti$_{2}$O$_{5}$ system. The frequency and Raman intensity of the related phonon modes are displayed in the table \ref{DFT-PBE-PhononFrequency}}
\label{Phonon23-30}
\end{figure*}
\begin{figure*}[t]
\includegraphics[width=16cm, trim = 0.6cm 0.5cm 0.3cm 0cm, clip]{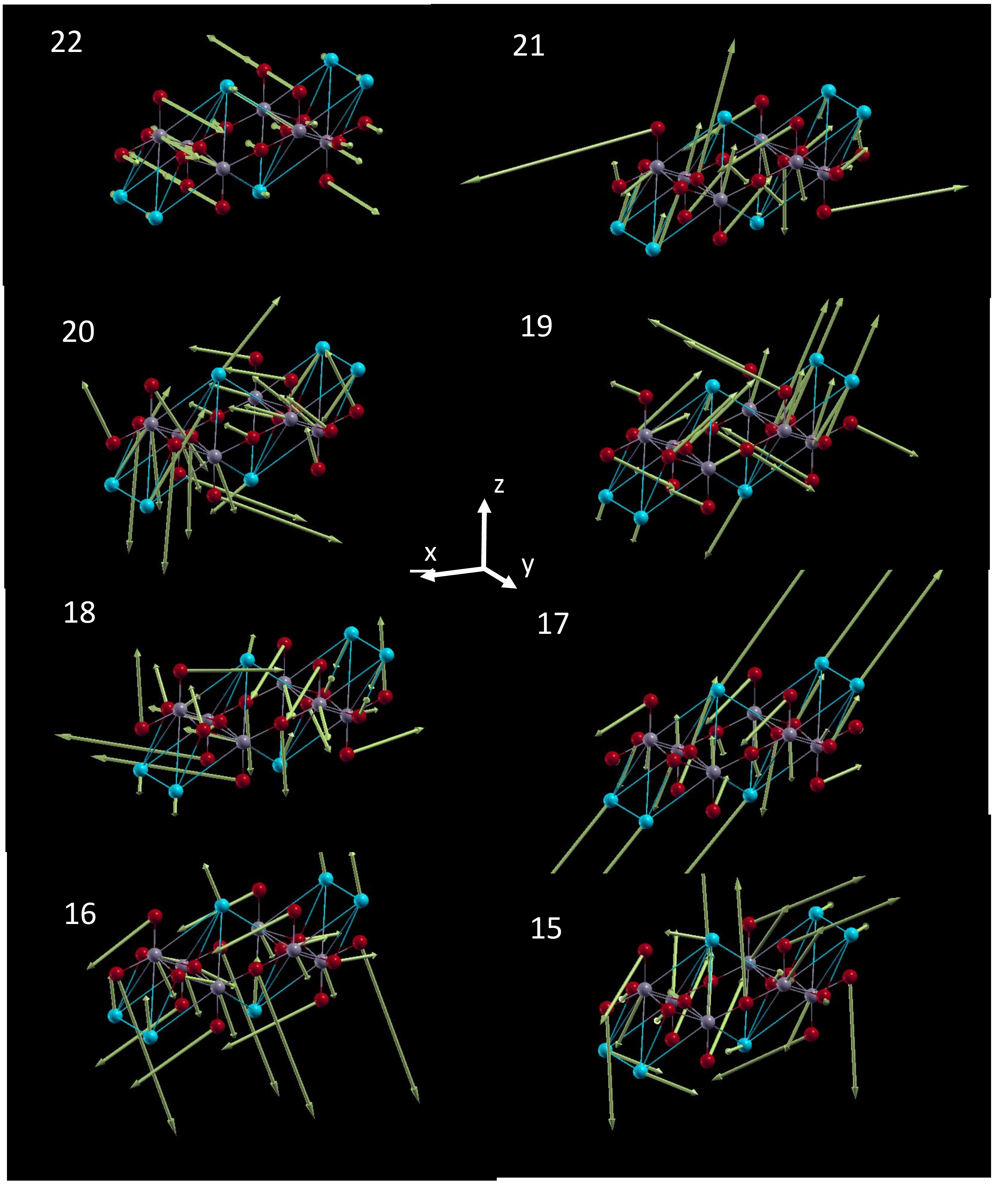}
\caption{Representations of the vibrational phonon modes N$\degree$ 22, 21, 20, 19, 18, 17, 16 and 15 calculated using Density Function Theory for Rb$_{2}$Ti$_{2}$O$_{5}$ system. The frequency and Raman intensity of the related phonon modes are displayed in the table \ref{DFT-PBE-PhononFrequency}}
\label{Phonon15-22}
\end{figure*}
\begin{figure*}[t]
\includegraphics[width=16cm, trim = 0.7cm 0.5cm 1.5cm 0.5cm, clip]{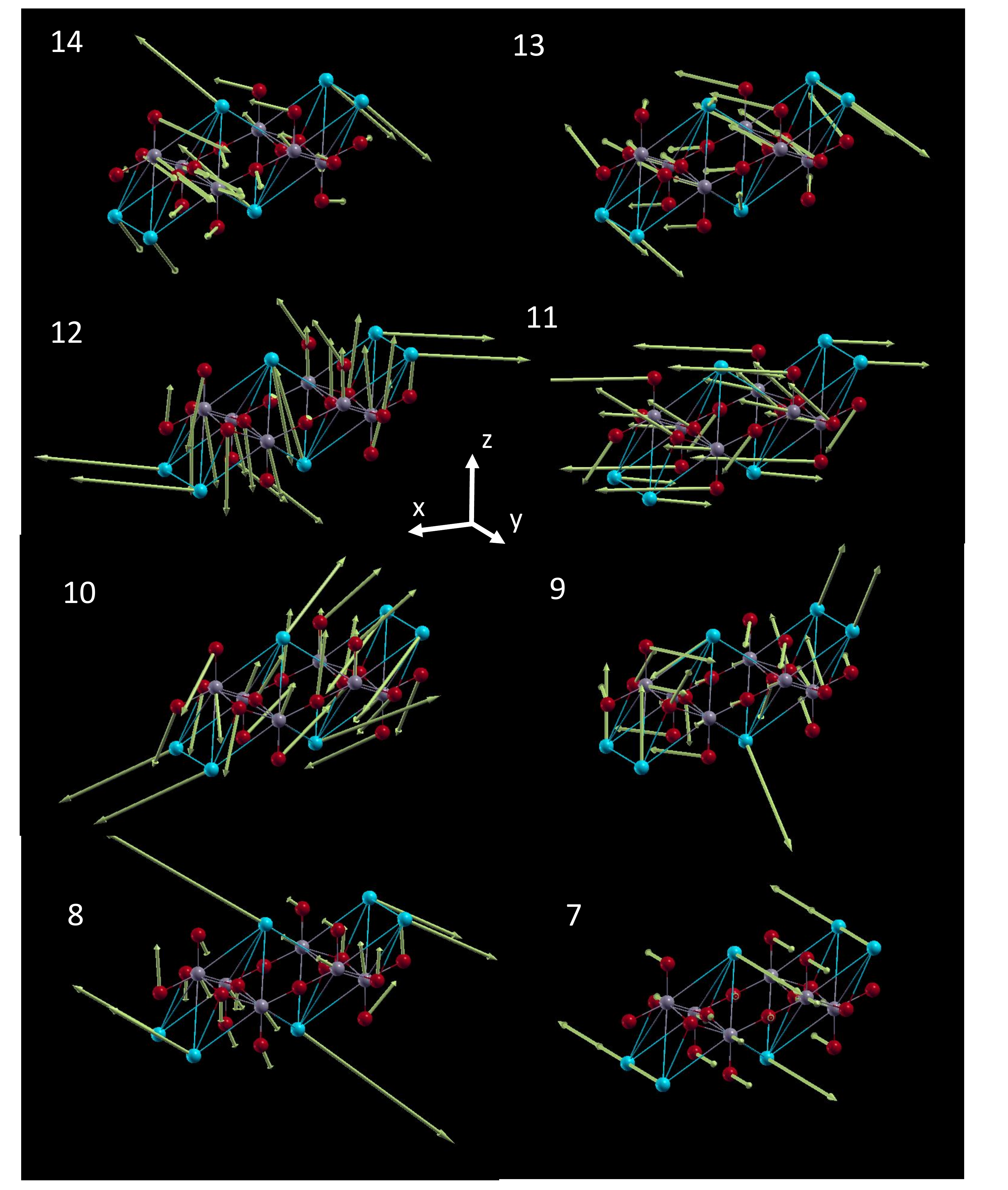}
\caption{Representations of the vibrational phonon modes N$\degree$ 14, 13, 12, 11, 10, 9, 8 and 7 calculated using Density Function Theory for Rb$_{2}$Ti$_{2}$O$_{5}$ system. The frequency and Raman intensity of the related phonon modes are displayed in the table \ref{DFT-PBE-PhononFrequency}}
\label{Phonon7-14}
\end{figure*}
\begin{figure*}[t]
\includegraphics[width=16cm, trim = 0.7cm 8cm 1.5cm 0.5cm, clip]{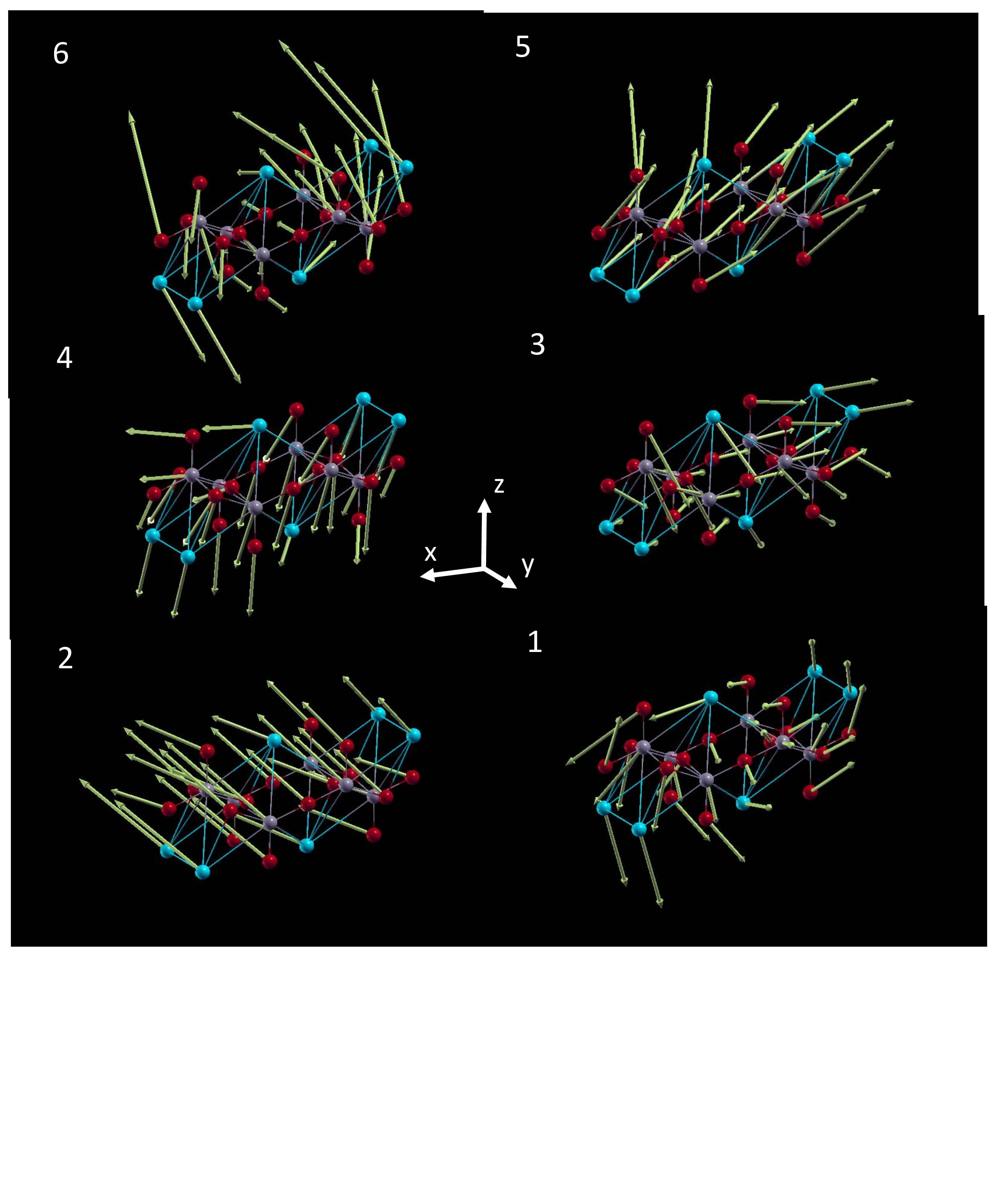}
\caption{Representations of the vibrational phonon modes N$\degree$ 6, 5, 4, 3, 2 and 1 calculated using Density Function Theory for Rb$_{2}$Ti$_{2}$O$_{5}$ system. The frequency and Raman intensity of the related phonon modes are displayed in the table \ref{DFT-PBE-PhononFrequency}}
\label{Phonon1-6}
\end{figure*}

Phonon mode calculations are based on the density functional Perturbation Theory \cite{DFT}. These calculations were performed by using LDA or GGA functionals and carried out at the corresponding relaxed structures. The phonon frequencies at the center of the Brillouin zone were computed with ultrasoft pseudopotentials, while the Raman intensities were estimated by employing norm-conserving pseudopotentials on top of the phonon calculations, and are therefore less accurate than the frequencies themselves. The theoretical phonon spectra are convolved with Lorentzians and plotted in figure \ref{RTO-Raman-300K}. The Raman intensities and the phonon frequencies are extracted from the DFT calculations whereas the full width at half maximum are arbitrary chosen as 1 to make them discernable. \newline
The red and blue curves represent the theoretical Raman spectra simulated with the GGA or the LDA functionals, respectively. The black square curve represents the experimental data recorded on a crystal of Rb$_2$Ti$_2$O$_5$ at room conditions between 180\,cm$^{-1}$ and 1300\,cm$^{-1}$. With respect to the experimental data, the computed frequencies are higher in the LDA and lower in the GGA. This can be linked to the fact that the LDA generally underestimates the lattice parameters and provides stiffer force constants, while the GGA makes the reverse. When looking at the high-frequency part of the spectrum, the four phonon modes are bracketed by the theoretical phonon modes found with GGA and LDA functionals, which differ by not more than 56\,cm$^{-1}$. These four phonon modes are denoted $\nu 1$, $\nu 2$, $\nu 3$ and $\nu 4$, from small to high frequencies. The representations of these four phonon modes are sketched in figure \ref{PhononModes}. They all correspond to vibration modes of Ti-O bonds. \newline
At low frequencies, ten theoretical modes were computed between 80\,cm$^{-1}$ and 330\,cm$^{-1}$. They can be unambigously linked to the set of experimental modes in the same wavenumber range, eventhough their intensities do not match with the experimental ones. The representations of all the calculated phonon modes are displayed in the supplemental material, in figures \ref{Phonon47-54}, \ref{Phonon39-46},  \ref{Phonon31-38}, \ref{Phonon23-30}, \ref{Phonon15-22}, \ref{Phonon7-14} and \ref{Phonon1-6}. \newline
Overall, the good agreement between the calculated phonon spectra for the simulated RTO system and the Raman spectra taken on a RTO crystal confirms that each mode can be assigned and analyzed in terms of a consistent set of atomic displacements and excludes the presence of singularities. \newline
The computed diagonal effective charges for Ti are $Z^{\ast}_{xx}=4.71, Z^{\ast}_{yy}=4.72, Z^{\ast}_{zz}=3.08$, with much smaller off-diagonal elements $Z^{\ast}_{xz}$ and $Z^{\ast}_{zx}$.
In contradiction to typical ferroelectric crystals such as BaTiO$_3$ \cite{Ghosez2007}, the effective charges do not show a relevant anomalous contribution, which is consistent with the absence of an electric-driven instability in Rb$_2$Ti$_2$O$_5$. 

\section{Relation with electrical properties}

\begin{figure}[h!]
\begin{center}
\includegraphics[width=8.3cm, trim = 0cm 0cm 0cm 0cm, clip]{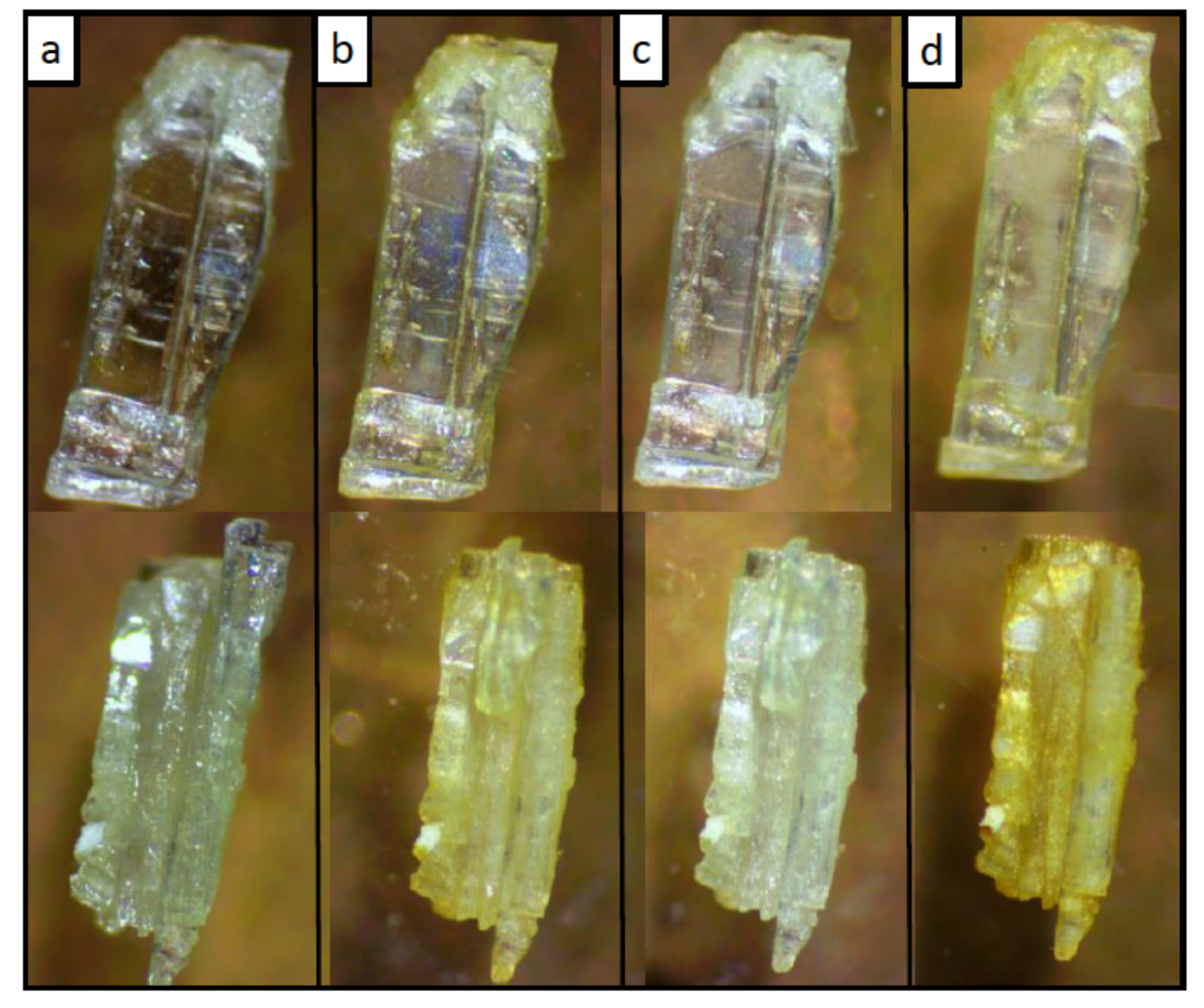}
\caption{Change of color observed when the material is alternatively annealed for two hours under vacuum and oxygen at 400\,K.}
\label{ColorChange}
\end{center}
\end{figure}

As reported earlier \cite{1-2016-RTO-Ionic}, the remarkable electrical properties of RTO are strongly enhanced for samples annealed under vacuum at 400\,K for a couple of hours. As a matter of fact, these electrical properties can also be "desactivated" when the material is annealed under 1\,bar-oxygen atmosphere at 400\,K. \newline
This observation tends to indicate that the existence of such electrical properties is corroborated with the presence of oxygen vacancies in the material. In addition, the figure \ref{ColorChange} shows that a reversible change of color operates when the material is alternatively annealed under vacuum and oxygen at 400\,K. This is a strong proof of the creation of oxygen vacancies since the material turns from white to yellow and yellow to white, similarly to what is known in TiO$_2$ \cite{Sekiya2000}. \newline

\begin{figure}[h!]
\begin{center}
\includegraphics[width=8.3cm, trim = 2cm 2cm 2cm 2cm, clip]{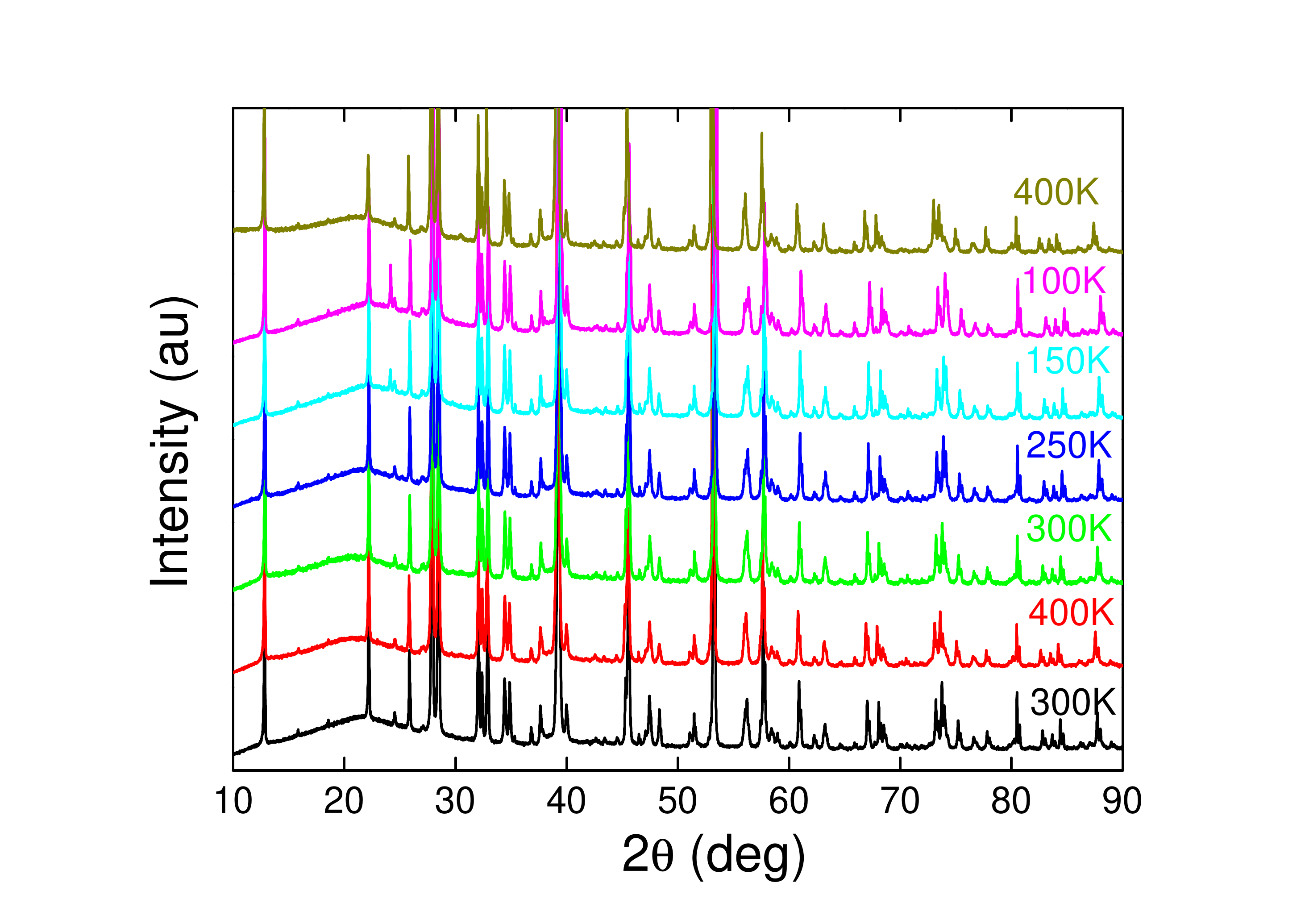}
\caption{Powder X-ray diffractograms of Rb$_{2}$Ti$_{2}$O$_{5}$ and Rb$_{2}$Ti$_{2}$O$_{5-\delta}$ taken at different temperatures and displaced for clarity. From bottom to top: as-grown sample at 300\,K, vacuum-annealed sample at 400\,K, vacuum-annealed sample at 300\,K, vacuum-annealed sample at 250\,K, vacuum-annealed sample at 150\,K, vacuum-annealed sample at 100\,K, N$_2$-annealed sample at 400\,K. All annealing procedures were made \textit{in situ}. The annealing duration was 8 hours.  The pic at 24.17 deg (3.68 angstroms) comes from the formation of a very thin layer of H$_2$O ice on the surface of the sample. Heating the sample from 150K up to 250K under vacuum allowed us to reversibly remove the ice. This proves that the sample is not altered by the presence of this residual solid state H$_2$O.}
\label{DRXinsitu}
\end{center}
\end{figure}

DRX investigation has thus been performed on RTO powder that was annealed \textit{in situ} in the diffraction chamber under N$_2$ atmosphere. The corresponding diffractogram is displayed in figure \ref{DRXinsitu} and does not show any changes as function of the temperature excepted for the apparition of a diffraction peak around 24\,degree below 250\,K, artifact due to a small contamination with ice into the diffraction chamber.  \newline
Powder diffraction at room temperature was also performed on samples annealed for 24\,hours at 400\,K under vacuum and again, the same crystal structure as in as-grown crystals was found.

The powder diffraction realized on as grown samples and displayed in  Tab \ref{PowderData} shows that the lattice parameters are not decreasing/increasing monotonously as expected as function of the temperature because of the thermal dilatations. The a lattice parameters decreases from 400\,K to 300\,K and then increases untill 100\,K, while b does the contrary and c decreases regularly. This behavior may be due to an unusual lattice distorsion which takes place roughly in the temperature range where transport and dielectric properties of the RTO are the most singular.  This would deserve further investigations. \newline
Finally, it is known that X-ray diffraction can not discern the presence of oxygen vacancies into a crystal structure. Neutron diffraction will be considered in order to further investigate the presence of oxygen vacancies induced by the annealing processes.\newline

\section{Conclusion}

The crystal structure of Rb$_{2}$Ti$_{2}$O$_{5}$ has been characterized as a function of temperature using X-Ray diffraction and Raman spectroscopy. Both methods found the space group of the material consistent with \textit{C}/2\textit{m} between 90\,K and 450\,K. The outcomes of the X-Ray diffraction and Raman spectroscopy are compared to theoretical calculations made with density functional theory and are in excellent agreement. \newline
The structure found is also confirmed by transmission electron microscopy and scanning electron microscopy. \newline
In addition,  this work shows that the \textit{C}2/\textit{m} structure of the material remains unchanged between 90\,K and 450\,K, thus excluding the existence of a structural transition between 200\,K and 330\,K. This is true for as-grown samples as well as in "deoxygenated" vacuum-annealed or N$_2$-annealed samples. 
These findings lead to discard a conventional ferroelectric transition as a mechanism responsible for the reported colossal polarisability of Rb$_{2}$Ti$_{2}$O$_{5}$. \newline

\bibliographystyle{apsrev4-1}

\begin{thebibliography}{10}%
\makeatletter
\providecommand \@ifxundefined [1]{%
 \ifx #1\undefined \expandafter \@firstoftwo
 \else \expandafter \@secondoftwo
\fi
}%
\providecommand \@ifnum [1]{%
 \ifnum #1\expandafter \@firstoftwo
 \else \expandafter \@secondoftwo
\fi
}%
\providecommand \enquote [1]{``#1''}%
\providecommand \bibnamefont  [1]{#1}%
\providecommand \bibfnamefont [1]{#1}%
\providecommand \citenamefont [1]{#1}%
\providecommand\href[0]{\@sanitize\@href}%
\providecommand\@href[1]{\endgroup\@@startlink{#1}\endgroup\@@href}%
\providecommand\@@href[1]{#1\@@endlink}%
\providecommand \@sanitize [0]{\begingroup\catcode`\&12\catcode`\#12\relax}%
\@ifxundefined \pdfoutput {\@firstoftwo}{%
 \@ifnum{\z@=\pdfoutput}{\@firstoftwo}{\@secondoftwo}%
}{%
 \providecommand\@@startlink[1]{\leavevmode}%
 \providecommand\@@endlink[0]{}%
}{%
 \providecommand\@@startlink[1]{%
  \leavevmode
  \pdfstartlink
   attr{/Border[0 0 1 ]/H/I/C[0 1 1]}%
   user{/Subtype/Link/A<</Type/Action/S/URI/URI(#1)>>}%
  \relax
 }%
 \providecommand\@@endlink[0]{\pdfendlink}%
}%
\providecommand \url  [0]{\begingroup\@sanitize \@url }%
\providecommand \@url [1]{\endgroup\@href {#1}{\urlprefix}}%
\providecommand \urlprefix [0]{URL }%
\providecommand \Eprint[0]{\href }%
\@ifxundefined \urlstyle {%
  \providecommand \doi [1]{doi:\discretionary{}{}{}#1}%
}{%
  \providecommand \doi [0]{doi:\discretionary{}{}{}\begingroup
  \urlstyle{rm}\Url }%
}%
\providecommand \doibase [0]{http://dx.doi.org/}%
\providecommand \Doi[1]{\href{\doibase#1}}%
\providecommand \bibAnnote [3]{%
  \BibitemShut{#1}%
  \begin{quotation}\noindent
    \textsc{Key:}\ #2\\\textsc{Annotation:}\ #3%
  \end{quotation}%
}%
\providecommand \bibAnnoteFile [2]{%
  \IfFileExists{#2}{\bibAnnote {#1} {#2} {\input{#2}}}{}%
}%
\providecommand \typeout [0]{\immediate \write \m@ne }%
\providecommand \selectlanguage [0]{\@gobble}%
\providecommand \bibinfo [0]{\@secondoftwo}%
\providecommand \bibfield [0]{\@secondoftwo}%
\providecommand \translation [1]{[#1]}%
\providecommand \BibitemOpen[0]{}%
\providecommand \bibitemStop [0]{}%
\providecommand \bibitemNoStop [0]{.\EOS\space}%
\providecommand \EOS [0]{\spacefactor3000\relax}%
\providecommand \BibitemShut [1]{\csname bibitem#1\endcsname}%
\bibitem{Na1.7K0.3Ti3O7}%
  \BibitemOpen
  \bibfield{author}{%
  \bibinfo {author} {\bibfnamefont{D.}~\bibnamefont{Maurya}}, \bibinfo {author}
  {\bibfnamefont{J.}~\bibnamefont{Kumar}},\ and\ \bibinfo {author}
  {\bibnamefont{Shripal}},\ }%
  \bibfield{journal}{%
  \bibinfo {journal} {Journal of Applied Physics}\ }%
  \textbf{\bibinfo {volume} {100}},\ \bibinfo {eid} {034103} (\bibinfo {year}
  {2006}),\ \doi{\bibinfo {doi} {http://dx.doi.org/10.1063/1.2227255}},\
  \url{http://scitation.aip.org/content/aip/journal/jap/100/3/10.1063/1.222725%
5}%
  \bibAnnoteFile{NoStop}{Na1.7K0.3Ti3O7}%
\bibitem{K2Ti6O13-1}%
  \BibitemOpen
  \bibfield{author}{%
  \bibinfo {author} {\bibfnamefont{S.~V.}\ \bibnamefont{Vikram}}, \bibinfo
  {author} {\bibfnamefont{D.~M.}\ \bibnamefont{Phase}},\ and\ \bibinfo {author}
  {\bibfnamefont{V.~S.}\ \bibnamefont{Chandel}},\ }%
  \bibfield{journal}{%
  \Doi{10.1007/s10854-009-0015-0}{\bibinfo {journal} {Journal of Materials
  Science: Materials in Electronics}}\ }%
  \textbf{\bibinfo {volume} {21}},\ \bibinfo {pages} {902} (\bibinfo {year}
  {2010}),\ ISSN \bibinfo {issn} {1573-482X},\
  \url{http://dx.doi.org/10.1007/s10854-009-0015-0}%
  \bibAnnoteFile{NoStop}{K2Ti6O13-1}%
\bibitem{Pyro-RTPO-2006}%
  \BibitemOpen
  \bibfield{author}{%
  \bibinfo {author} {\bibfnamefont{Y.}~\bibnamefont{Shaldin}}, \bibinfo
  {author} {\bibfnamefont{S.}~\bibnamefont{Matyjasik}}, \bibinfo {author}
  {\bibfnamefont{M.}~\bibnamefont{Tseitlin}},\ and\ \bibinfo {author}
  {\bibfnamefont{M.}~\bibnamefont{Roth}},\ }%
  \bibfield{journal}{%
  \Doi{http://dx.doi.org/10.1016/j.optmat.2006.11.005}{\bibinfo {journal}
  {Optical Materials}}\ }%
  \textbf{\bibinfo {volume} {30}},\ \bibinfo {pages} {101 } (\bibinfo {year}
  {2007}),\ ISSN \bibinfo {issn} {0925-3467},\
  \url{http://www.sciencedirect.com/science/article/pii/S0925346706003703}%
  \bibAnnoteFile{NoStop}{Pyro-RTPO-2006}%
\bibitem{Pyro-KTAO-2010}%
  \BibitemOpen
  \bibfield{author}{%
  \bibinfo {author} {\bibfnamefont{Y.~V.}\ \bibnamefont{Shaldin}}, \bibinfo
  {author} {\bibfnamefont{S.}~\bibnamefont{Matyjasik}}, \bibinfo {author}
  {\bibfnamefont{M.}~\bibnamefont{Tseitlin}}, \bibinfo {author}
  {\bibfnamefont{E.}~\bibnamefont{Mojaev}},\ and\ \bibinfo {author}
  {\bibfnamefont{M.}~\bibnamefont{Roth}},\ }%
  \bibfield{journal}{%
  \Doi{10.1002/pssb.200844294}{\bibinfo {journal} {physica status solidi (b)}}\
  }%
  \textbf{\bibinfo {volume} {246}},\ \bibinfo {pages} {452} (\bibinfo {year}
  {2009}),\ ISSN \bibinfo {issn} {1521-3951},\
  \url{http://dx.doi.org/10.1002/pssb.200844294}%
  \bibAnnoteFile{NoStop}{Pyro-KTAO-2010}%
\bibitem{CMR}%
  \BibitemOpen
  \bibfield{author}{%
  \bibinfo {author} {\bibfnamefont{M.}~\bibnamefont{McCormack}}, \bibinfo
  {author} {\bibfnamefont{S.}~\bibnamefont{Jin}}, \bibinfo {author}
  {\bibfnamefont{T.~H.}\ \bibnamefont{Tiefel}}, \bibinfo {author}
  {\bibfnamefont{R.~M.}\ \bibnamefont{Fleming}}, \bibinfo {author}
  {\bibfnamefont{J.~M.}\ \bibnamefont{Phillips}},\ and\ \bibinfo {author}
  {\bibfnamefont{R.}~\bibnamefont{Ramesh}},\ }%
  \bibfield{journal}{%
  \Doi{http://dx.doi.org/10.1063/1.111372}{\bibinfo {journal} {Applied Physics
  Letters}}\ }%
  \textbf{\bibinfo {volume} {64}},\ \bibinfo {pages} {3045} (\bibinfo {year}
  {1994}),\
  \url{http://scitation.aip.org/content/aip/journal/apl/64/22/10.1063/1.111372%
}%
  \bibAnnoteFile{NoStop}{CMR}%
\bibitem{LAOSTO}%
  \BibitemOpen
  \bibfield{author}{%
  \bibinfo {author} {\bibfnamefont{N.}~\bibnamefont{Reyren}}, \bibinfo {author}
  {\bibfnamefont{S.}~\bibnamefont{Thiel}}, \bibinfo {author}
  {\bibfnamefont{A.~D.}\ \bibnamefont{Caviglia}}, \bibinfo {author}
  {\bibfnamefont{L.~F.}\ \bibnamefont{Kourkoutis}}, \bibinfo {author}
  {\bibfnamefont{G.}~\bibnamefont{Hammerl}}, \bibinfo {author}
  {\bibfnamefont{C.}~\bibnamefont{Richter}}, \bibinfo {author}
  {\bibfnamefont{C.~W.}\ \bibnamefont{Schneider}}, \bibinfo {author}
  {\bibfnamefont{T.}~\bibnamefont{Kopp}}, \bibinfo {author}
  {\bibfnamefont{A.-S.}\ \bibnamefont{Rüetschi}}, \bibinfo {author}
  {\bibfnamefont{D.}~\bibnamefont{Jaccard}}, \bibinfo {author}
  {\bibfnamefont{M.}~\bibnamefont{Gabay}}, \bibinfo {author}
  {\bibfnamefont{D.~A.}\ \bibnamefont{Muller}}, \bibinfo {author}
  {\bibfnamefont{J.-M.}\ \bibnamefont{Triscone}},\ and\ \bibinfo {author}
  {\bibfnamefont{J.}~\bibnamefont{Mannhart}},\ }%
  \bibfield{journal}{%
  \bibinfo {journal} {Science}\ }%
  \textbf{\bibinfo {volume} {317}},\ \bibinfo {pages} {1196} (\bibinfo {year}
  {2007}),\
  \url{http://science.sciencemag.org/content/317/5842/1196.full.pdf+html}%
  \bibAnnoteFile{NoStop}{LAOSTO}%
\bibitem{K2Ti2O5-1986}%
  \BibitemOpen
  \bibfield{author}{%
  \bibinfo {author} {\bibfnamefont{M.}~\bibnamefont{Tournoux}}, \bibinfo
  {author} {\bibfnamefont{R.}~\bibnamefont{Marchand}},\ and\ \bibinfo {author}
  {\bibfnamefont{L.}~\bibnamefont{Brohan}},\ }%
  \bibfield{journal}{%
  \Doi{http://dx.doi.org/10.1016/0079-6786(86)90003-8}{\bibinfo {journal}
  {Progress in Solid State Chemistry}}\ }%
  \textbf{\bibinfo {volume} {17}},\ \bibinfo {pages} {33 } (\bibinfo {year}
  {1986}),\ ISSN \bibinfo {issn} {0079-6786},\
  \url{http://www.sciencedirect.com/science/article/pii/0079678686900038}%
  \bibAnnoteFile{NoStop}{K2Ti2O5-1986}%
\bibitem{Shripal2005}%
  \BibitemOpen
  \bibfield{author}{%
  \bibinfo {author} {\bibnamefont{Shripal}}, \bibinfo {author}
  {\bibfnamefont{S.}~\bibnamefont{Badhwar}}, \bibinfo {author}
  {\bibfnamefont{D.}~\bibnamefont{Maurya}},\ and\ \bibinfo {author}
  {\bibfnamefont{J.}~\bibnamefont{Kumar}},\ }%
  \bibfield{journal}{%
  \Doi{10.1007/s10854-005-2723-4}{\bibinfo {journal} {Journal of Materials
  Science: Materials in Electronics}}\ }%
  \textbf{\bibinfo {volume} {16}},\ \bibinfo {pages} {495} (\bibinfo {year}
  {2005}),\ ISSN \bibinfo {issn} {1573-482X},\
  \url{http://dx.doi.org/10.1007/s10854-005-2723-4}%
  \bibAnnoteFile{NoStop}{Shripal2005}%
\bibitem{Rb2Ti2O5-1}%
  \BibitemOpen
  \bibfield{author}{%
  \bibinfo {author} {\bibfnamefont{O.}~\bibnamefont{Schmitz-Dumont}}\ and\
  \bibinfo {author} {\bibfnamefont{A.~H.}\ \bibnamefont{Schulz}},\ }%
  \bibfield{journal}{%
  \Doi{10.1007/BF00897713}{\bibinfo {journal} {Monatshefte f{\"u}r Chemie und
  verwandte Teile anderer Wissenschaften}}\ }%
  \textbf{\bibinfo {volume} {83}},\ \bibinfo {pages} {638} (\bibinfo {year}
  {1952}),\ ISSN \bibinfo {issn} {1434-4475},\
  \url{http://dx.doi.org/10.1007/BF00897713}%
  \bibAnnoteFile{NoStop}{Rb2Ti2O5-1}%
\bibitem{Rb2Ti2O5-2}%
  \BibitemOpen
  \bibfield{author}{%
  \bibinfo {author} {\bibfnamefont{O.}~\bibnamefont{Schmitz-DuMont}}\ and\
  \bibinfo {author} {\bibfnamefont{H.}~\bibnamefont{Reckhard}},\ }%
  \bibfield{journal}{%
  \Doi{10.1007/BF00925227}{\bibinfo {journal} {Monatshefte f{\"u}r Chemie und
  verwandte Teile anderer Wissenschaften}}\ }%
  \textbf{\bibinfo {volume} {90}},\ \bibinfo {pages} {134} (\bibinfo {year}
  {1959}),\ ISSN \bibinfo {issn} {1434-4475},\
  \url{http://dx.doi.org/10.1007/BF00925227}%
  \bibAnnoteFile{NoStop}{Rb2Ti2O5-2}%
\bibitem{K2Ti2O5-1960}%
  \BibitemOpen
  \bibfield{author}{%
  \bibinfo {author} {\bibfnamefont{S.}~\bibnamefont{Andersson}}\ and\ \bibinfo
  {author} {\bibfnamefont{A.~D.}\ \bibnamefont{Wadsley}},\ }%
  \bibfield{journal}{%
  \bibinfo {journal} {Nature}\ }%
  \textbf{\bibinfo {volume} {4736}} (\bibinfo {month} {aug}\ \bibinfo {year}
  {1960}),\ \doi{\bibinfo {doi} {10.1038/187499a0}},\
  \url{http://www.nature.com/nature/journal/v187/n4736/pdf/187499a0.pdf}%
  \bibAnnoteFile{NoStop}{K2Ti2O5-1960}%
\bibitem{K2Ti2O5-1961}%
  \BibitemOpen
  \bibfield{author}{%
  \bibinfo {author} {\bibfnamefont{S.}~\bibnamefont{Andersson}}\ and\ \bibinfo
  {author} {\bibfnamefont{A.~D.}\ \bibnamefont{Wadsley}},\ }%
  \bibfield{journal}{%
  \bibinfo {journal} {Acta Chem. Scand}\ }%
  \textbf{\bibinfo {volume} {3}} (\bibinfo {month} {aug}\ \bibinfo {year}
  {1961}),\ \doi{\bibinfo {doi} {10.3891/acta.chem.scand.15-0663}},\
  \url{http://actachemscand.org/pdf/acta_vol_15_p0663-0669.pdf}%
  \bibAnnoteFile{NoStop}{K2Ti2O5-1961}%
\bibitem{K2Ti2O5-2009}%
  \BibitemOpen
  \bibfield{author}{%
  \bibinfo {author} {\bibfnamefont{Z.~G.}\ \bibnamefont{Qiang~Wang}}\ and\
  \bibinfo {author} {\bibfnamefont{J.~S.}\ \bibnamefont{Chung}},\ }%
  \bibfield{journal}{%
  \Doi{10.1039/B909455E}{\bibinfo {journal} {Chem. Commun}}\ }%
  \textbf{\bibinfo {volume} {35}},\ \bibinfo {pages} {5284} (\bibinfo {year}
  {2009}),\ \url{http://pubs.rsc.org/en/content/articlepdf/2009/cc/b909455e}%
  \bibAnnoteFile{NoStop}{K2Ti2O5-2009}%
\bibitem{K2Ti2O5-2011}%
  \BibitemOpen
  \bibfield{author}{%
  \bibinfo {author} {\bibfnamefont{J.~S.~C.}\ \bibnamefont{Qiang~Wang}}\ and\
  \bibinfo {author} {\bibfnamefont{Z.}~\bibnamefont{Guo}},\ }%
  \bibfield{journal}{%
  \Doi{10.1021/ie200698j}{\bibinfo {journal} {Ind. Eng. Chem. Res}}\ }%
  \textbf{\bibinfo {volume} {50 (13)}},\ \bibinfo {pages} {8384} (\bibinfo
  {year} {2011}),\ \url{http://pubs.acs.org/doi/abs/10.1021/ie200698j}%
  \bibAnnoteFile{NoStop}{K2Ti2O5-2011}%
\bibitem{1-2016-RTO-Ionic}%
  \BibitemOpen
  \bibfield{author}{%
  \bibinfo {author} {\bibfnamefont{R.}~\bibnamefont{Federicci}}, \bibinfo
  {author} {\bibfnamefont{S.}~\bibnamefont{Holé}}, \bibinfo {author}
  {\bibfnamefont{B.}~\bibnamefont{Baptiste}}, \bibinfo {author}
  {\bibfnamefont{S.}~\bibnamefont{Mercone}},\ and\ \bibinfo {author}
  {\bibfnamefont{B.}~\bibnamefont{léridon}}}%
   (\bibinfo {year} {2017}),\ \url{http://}%
  \bibAnnoteFile{NoStop}{1-2016-RTO-Ionic}%
\bibitem{DRX1}%
  \BibitemOpen
  \bibfield{journal}{%
  \bibinfo {journal} {Rigaku Oxford Diffraction}}%
   (\bibinfo {year} {2015})%
  \bibAnnoteFile{NoStop}{DRX1}%
\bibitem{DRX4}%
  \BibitemOpen
  \bibfield{author}{%
  \bibinfo {author} {\bibfnamefont{H.~M.}\ \bibnamefont{Rietveld}},\ }%
  \bibfield{journal}{%
  \bibinfo {journal} {J. Appl. Crystallogrr}\ }%
  \textbf{\bibinfo {volume} {65–71}} (\bibinfo {year} {1969})%
  \bibAnnoteFile{NoStop}{DRX4}%
\bibitem{DRX5}%
  \BibitemOpen
  \bibfield{author}{%
  \bibinfo {author} {\bibnamefont{Rodriguez-Carvajal}},\ }%
  \bibfield{journal}{%
  \bibinfo {journal} {J. Phys. B}\ }%
  \textbf{\bibinfo {volume} {55–69}} (\bibinfo {year} {1993})%
  \bibAnnoteFile{NoStop}{DRX5}%
\bibitem{VESTA}%
  \BibitemOpen
  \bibfield{author}{%
  \bibinfo {author} {\bibfnamefont{K.}~\bibnamefont{Momma}}\ and\ \bibinfo
  {author} {\bibfnamefont{F.}~\bibnamefont{Izumi}},\ }%
  \bibfield{journal}{%
  \Doi{10.1107/S0021889811038970}{\bibinfo {journal} {Journal of Applied
  Crystallography}}\ }%
  \textbf{\bibinfo {volume} {44}},\ \bibinfo {pages} {1272} (\bibinfo {month}
  {Dec}\ \bibinfo {year} {2011}),\
  \url{https://doi.org/10.1107/S0021889811038970}%
  \bibAnnoteFile{NoStop}{VESTA}%
\bibitem{QE}%
  \BibitemOpen
  \bibfield{author}{%
  \bibinfo {author} {\bibfnamefont{e.~a.}\ \bibnamefont{Paolo~Giannozzi}},\ }%
  \bibfield{journal}{%
  \bibinfo {journal} {Journal of Physics: Condensed Matter}\ }%
  \textbf{\bibinfo {volume} {21}},\ \bibinfo {pages} {395502} (\bibinfo {year}
  {2009}),\ \url{http://stacks.iop.org/0953-8984/21/i=39/a=395502}%
  \bibAnnoteFile{NoStop}{QE}%
\bibitem{PBE}%
  \BibitemOpen
  \bibfield{author}{%
  \bibinfo {author} {\bibfnamefont{J.~P.}\ \bibnamefont{Perdew}}, \bibinfo
  {author} {\bibfnamefont{K.}~\bibnamefont{Burke}},\ and\ \bibinfo {author}
  {\bibfnamefont{M.}~\bibnamefont{Ernzerhof}},\ }%
  \bibfield{journal}{%
  \Doi{10.1103/PhysRevLett.77.3865}{\bibinfo {journal} {Phys. Rev. Lett.}}\ }%
  \textbf{\bibinfo {volume} {77}},\ \bibinfo {pages} {3865} (\bibinfo {month}
  {Oct}\ \bibinfo {year} {1996}),\
  \url{http://link.aps.org/doi/10.1103/PhysRevLett.77.3865}%
  \bibAnnoteFile{NoStop}{PBE}%
\bibitem{KohmSham1}%
  \BibitemOpen
  \bibfield{author}{%
  \bibinfo {author} {\bibfnamefont{P.}~\bibnamefont{Hohenberg}}\ and\ \bibinfo
  {author} {\bibfnamefont{W.}~\bibnamefont{Kohn}},\ }%
  \bibfield{journal}{%
  \Doi{10.1103/PhysRev.136.B864}{\bibinfo {journal} {Phys. Rev.}}\ }%
  \textbf{\bibinfo {volume} {136}},\ \bibinfo {pages} {B864} (\bibinfo {month}
  {Nov}\ \bibinfo {year} {1964}),\
  \url{http://link.aps.org/doi/10.1103/PhysRev.136.B864}%
  \bibAnnoteFile{NoStop}{KohmSham1}%
\bibitem{KohmSham2}%
  \BibitemOpen
  \bibfield{author}{%
  \bibinfo {author} {\bibfnamefont{W.}~\bibnamefont{Kohn}}\ and\ \bibinfo
  {author} {\bibfnamefont{L.~J.}\ \bibnamefont{Sham}},\ }%
  \bibfield{journal}{%
  \Doi{10.1103/PhysRev.140.A1133}{\bibinfo {journal} {Phys. Rev.}}\ }%
  \textbf{\bibinfo {volume} {140}},\ \bibinfo {pages} {A1133} (\bibinfo {month}
  {Nov}\ \bibinfo {year} {1965}),\
  \url{http://link.aps.org/doi/10.1103/PhysRev.140.A1133}%
  \bibAnnoteFile{NoStop}{KohmSham2}%
\bibitem{Murnaghan3}%
  \BibitemOpen
  \bibfield{author}{%
  \bibinfo {author} {\bibfnamefont{F.~D.}\ \bibnamefont{Murnaghan}},\ }%
  \bibfield{journal}{%
  \bibinfo {journal} {PNAS}\ }%
  \textbf{\bibinfo {volume} {30}},\ \bibinfo {pages} {244} (\bibinfo {year}
  {1944}),\ \url{http://www.pnas.org/content/30/9/244.citation}%
  \bibAnnoteFile{NoStop}{Murnaghan3}%
\bibitem{DFTPath}%
  \BibitemOpen
  \bibfield{author}{%
  \bibinfo {author} {\bibfnamefont{W.}~\bibnamefont{Setyawan}}\ and\ \bibinfo
  {author} {\bibfnamefont{S.}~\bibnamefont{Curtarolo}},\ }%
  \bibfield{journal}{%
  \Doi{http://dx.doi.org/10.1016/j.commatsci.2010.05.010}{\bibinfo {journal}
  {Computational Materials Science}}\ }%
  \textbf{\bibinfo {volume} {49}},\ \bibinfo {pages} {299 } (\bibinfo {year}
  {2010}),\ ISSN \bibinfo {issn} {0927-0256},\
  \url{http://www.sciencedirect.com/science/article/pii/S0927025610002697}%
  \bibAnnoteFile{NoStop}{DFTPath}%
  
\bibitem{DFT}%
  \BibitemOpen
  \bibfield{author}{%
  \bibinfo {author} {\bibfnamefont{S.}~\bibnamefont{Baroni}}, \bibinfo {author}
  {\bibfnamefont{S.}~\bibnamefont{de~Gironcoli}}, \bibinfo {author}
  {\bibfnamefont{A.}~\bibnamefont{Dal~Corso}},\ and\ \bibinfo {author}
  {\bibfnamefont{P.}~\bibnamefont{Giannozzi}},\ }%
  \bibfield{journal}{%
  \Doi{10.1103/RevModPhys.73.515}{\bibinfo {journal} {Rev. Mod. Phys.}}\ }%
  \textbf{\bibinfo {volume} {73}},\ \bibinfo {pages} {515} (\bibinfo {month}
  {Jul}\ \bibinfo {year} {2001}),\
  \url{http://link.aps.org/doi/10.1103/RevModPhys.73.515}%
  \bibAnnoteFile{NoStop}{DFT}%
  
\bibitem{Ghosez2007}%
  \BibitemOpen
  \bibfield{author}{%
  \bibinfo {author} {\bibfnamefont{P.}~\bibnamefont{Rabe}, \bibfnamefont{Karin
  M.and~Ghosez}},\ }%
  \enquote{\bibinfo {title} {First-principles studies of
  ferroelectricoxides},}\ in\ \Doi{10.1007/978-3-540-34591-6_4}{\emph{\bibinfo
  {booktitle} {Physics of Ferroelectrics: A Modern Perspective}}}\ (\bibinfo
  {publisher} {Springer Berlin Heidelberg},\ \bibinfo {address} {Berlin,
  Heidelberg},\ \bibinfo {year} {2007})\ pp.\ \bibinfo {pages} {117--174},\
  ISBN \bibinfo {isbn} {978-3-540-34591-6},\
  \url{http://dx.doi.org/10.1007/978-3-540-34591-6_4}%
  \bibAnnoteFile{NoStop}{Ghosez2007}%
  
  \bibitem{Sekiya2000} 
   \BibitemOpen
  \bibfield{author}{%
  \bibinfo {author} {T. Sekiya},   \bibinfo {author} {K. Ichimura},   \bibinfo {author} { M. Igarashi} and \bibinfo {author} { S. Kurita},\ }%
 \bibfield{journal}{%
  \Doi{http://dx.doi.org/10.1016/S0022-3697(99)00424-2}{\bibinfo {journal} {Journal of Physics and Chemistry of Solids}}\ }%
\textbf{\bibinfo {volume} {61}},\ \bibinfo {pages} {1237--1242} (\bibinfo {month}
  {Aug}\ \bibinfo {year} {2000}),\
 \url{ http://www.sciencedirect.com/science/article/pii/S0022369799004242}%
    \bibAnnoteFile{NoStop}{Sekiya2000}%


\end{thebibliography}

%

\newpage
\newpage

\onecolumngrid
\section{Appendix A : tables for different XRD results}

\begin{table}[!h]
\centering
\begin{tabular}{||c|c|c|c|c|c|c||}
\hline\textbf{Atom} & \textbf{U$_{11}$} & \textbf{U$_{22}$} & \textbf{U$_{33}$} & \textbf{U$_{23}$}&\textbf{ U$_{13}$} &\textbf{U$_{12}$}\\
\hline Rb1 & 18.9(4) & 21.0(4) & 24.1(4) & 0 & 6.2(2) & 0 \\
\hline Ti1 & 4.4(5) & 10.2(6) & 17.3(5) & 0 & 0.7(4) & 0 \\
\hline O1 & 9(3) & 13(3) & 40(3) & 0 & 9(2) & 0\\
\hline O2 & 19(2) & 22(3) & 18.6(19) & 0 & -1.1(16) & 0\\
\hline O3 & 5.7(16) & 12(2) & 24.0(19) & 0 & -1.7(14) & 0  \\
\hline
\end{tabular}
\caption{\textbf{Anisotropic Displacement Parameters (\AA $^2 \times$10$^3$) for Rb$_2$Ti$_2$O$_5$ at 300K. The anisotropic displacement factor exponent takes the form: -2$\pi^2(h^2a^{*2}U_{11}+2hka^*b^*U_{12}+)$.}}
\label{tabcrys3}
\end{table}

\begin{table}[!h]
\centering
\begin{tabular}{||c|c|c|c|c|c|c||}
\hline \textbf{Atom} & \textbf{Atom} & \textbf{Length(\AA)} &  & \textbf{Atom} & \textbf{Atom} & \textbf{Length(\AA)} \\
\hline Rb1 & Rb1$^1$ & 3.3468(11) &  & Ti1 & O3$^9$ & 1.9861(10)\\
\hline Rb1 & Ti1$^2$ & 3.5945(10) &  & Ti1 & O3 & 1.988(4)\\
\hline Rb1 & Ti1$^3$ & 3.5945(10) &  & Ti1 & O3$^{10}$ & 1.9861(10)\\
\hline Rb1 & O3  & 2.878(4) &  & Ti1 & O1 & 1.8842(9)\\
\hline Rb1 & O3$^4$ &3.537(4) &  & Ti1 & O2 & 1.709(4)\\ 
\hline Rb1 & O1$^2$ &3.1400(5) &  & O3 & Rb1$^4$ & 3.537(4)\\
\hline Rb1 & O1$^3$ &3.1400(5) &  & O3 & Ti1$^{10}$ & 1.9861(10)\\
\hline Rb1 & O2$^2$ &3.153(3) &  & O3 & Ti1$^{9}$ & 1.9861(10)\\
\hline Rb1 & O2$^5$ &2.894(3) &  & O1 & Rb1$^{9}$ & 3.1400(5)\\
\hline Rb1 & O2$^3$ &3.153(3)&  & O1 & Rb1$^{8}$ & 3.1400(5)\\
\hline Rb1 & O2 &3.164(4) &  & O1 & Rb1$^{10}$ &3.1400(5)\\
\hline Rb1 & O2$^6$ &2.894(3) &  & O1 & Rb1$^{7}$ & 3.1400(5)\\
\hline Ti1 & Rb1$^7$ & 3.5945(10) &  & O1 & Ti$^{11}$& 1.8841(9) \\
\hline Ti1 & Rb1$^8$ & 3.5945(10) &  & O2 & RbA$^{6}$& 2.894(3) \\
\hline Ti1 & Rb1$^9$ & 3.7281(10)&  & O2 & Rb1$^{7}$& 3.153(3) \\
\hline Ti1 & Rb1$^{10}$ & 3.7281(10) &  & O2 & Rb1$^{8}$&3.153(3)\\
\hline Ti1 & Ti1$^{10}$ & 3.0920(14) &  & O2 & Rb1$^{5}$& 2.894(3) \\
\hline Ti1 & Ti1$^{9}$ & 3.0920(14) &  &   &  &   \\
\hline
\end{tabular}
\caption{\textbf{Bond Lengths for  for Rb$_2$Ti$_2$O$_5$ at 300K. }\\$^1$-X,1-Y,2-Z; $^2$-1/2+X,-1/2+Y,+Z; $^3$-1/2+X,1/2+Y,+Z; $^4$-X,1-Y,1-Z; $^5$1/2-X,3/2-Y,2-Z; $^6$1/2-X,1/2-Y,2-Z; $^7$1/2+X,1/2+Y,+Z; $^8$1/2+X,-1/2+Y,+Z; $^9$1/2-X,1/2-Y,1-Z; $^{10}$1/2-X,3/2-Y,1-Z; $^{11}$1-X,1-Y,1-Z}
\label{tabcrys4}
\end{table}

\begin{table}[!h]
\centering
\begin{tabular}{||c|c|c|c|c|c|c||}
\hline\textbf{Atom} & \textbf{U$_{11}$} & \textbf{U$_{22}$} & \textbf{U$_{33}$} & \textbf{U$_{23}$}&\textbf{ U$_{13}$} &\textbf{U$_{12}$}\\
\hline Rb1 & 25.2(3) & 27.7(3) & 28.7(3) & 0 & 8.87(19) & 0 \\
\hline Ti1 & 5.8(3) & 12.6(4) & 19.2(4) & 0 & 1.6(2) & 0 \\
\hline O1 & 10(2) & 19(3) & 48(3) & 0 & 15(2) & 0\\
\hline O2 & 29.5(19) & 24(2) & 21.9(16) & 0 & -0.5(14) & 0\\
\hline O3 & 5.5(12) & 13.6(15) & 31.8(17) & 0 & 0.2(11) & 0  \\
\hline
\end{tabular}
\caption{\textbf{Anisotropic Displacement Parameters (\AA $^2 \times$10$^3$) for Rb$_2$Ti$_2$O$_5$ at 400K. The anisotropic displacement factor exponent takes the form: -2$\pi^2(h^2a^{*2}U_{11}+2hka^*b^*U_{12}+)$.}}
\label{tabcrys3}
\end{table}

\begin{table}[!h]
\centering
\begin{tabular}{||c|c|c|c|c|c|c||}
\hline \textbf{Atom} & \textbf{Atom} & \textbf{Length(\AA)} &  & \textbf{Atom} & \textbf{Atom} & \textbf{Length(\AA)} \\
\hline Rb1 & Rb1$^1$ & 3.3518(10) &  & Ti1 & O3$^9$ & 1.9824(9)\\
\hline Rb1 & Ti1$^2$ & 3.6000(9) &  & Ti1 & O3 & 1.992(3)\\
\hline Rb1 & Ti1$^3$ & 3.6000(9) &  & Ti1 & O3$^{10}$ & 1.9824(9)\\
\hline Rb1 & O3  & 2.893(4) &  & Ti1 & O1 & 1.8844(8)\\
\hline Rb1 & O3$^4$ &3.542(3) &  & Ti1 & O2 & 1.705(4)\\ 
\hline Rb1 & O1$^3$ &3.1464(5) &  & O3 & Rb1$^4$ & 3.542(3)\\
\hline Rb1 & O1$^2$ &3.1464(5) &  & O3 & Ti1$^{10}$ & 1.9824(9)\\
\hline Rb1 & O2$^2$ &3.154(3) &  & O3 & Ti1$^{9}$ & 1.9824(9)\\
\hline Rb1 & O2 & 3.164(4) &  & O1 & Rb1$^{7}$ & 3.1464(5)\\
\hline Rb1 & O2$^5$ &2.899(3)&  & O1 & Rb1$^{10}$ & 3.1464(5)\\
\hline Rb1 & O2$^3$ &3.154(3) &  & O1 & Rb1$^{8}$ & 3.1464(5)\\
\hline Rb1 & O2$^6$ &2.899(3) &  & O1 & Rb1$^{9}$ & 3.1464(5)\\
\hline Ti1 & Rb1$^7$ & 3.6000(9) &  & O1 & Ti$^{11}$& 1.8844(7) \\
\hline Ti1 & Rb1$^8$ & 3.6000(9) &  & O2 & Rb1$^{6}$& 2.899(3) \\
\hline Ti1 & Rb1$^9$ & 3.7338(9)&  & O2 & Rb1$^{8}$& 3.154(3) \\
\hline Ti1 & Rb1$^{10}$ & 3.7338(9) &  & O2 & Rb1$^{5}$& 2.899(3)\\
\hline Ti1 & Rb1$^{10}$ & 3.0926(12) &  & O2 & Rb1$^{7}$& 3.154(3)) \\
\hline Ti1 & Rb1$^{9}$ & 3.0926(12) &  &   &  &   \\
\hline
\end{tabular}
\caption{\textbf{Bond Lengths for  for Rb$_2$Ti$_2$O$_5$ at 400K. }\\$^1$-X,1-Y,2-Z; $^2$-1/2+X,-1/2+Y,+Z; $^3$-1/2+X,1/2+Y,+Z; $^4$-X,1-Y,1-Z; $^5$1/2-X,3/2-Y,2-Z; $^6$1/2-X,1/2-Y,2-Z; $^7$1/2+X,1/2+Y,+Z; $^8$1/2+X,-1/2+Y,+Z; $^9$1/2-X,1/2-Y,1-Z; $^{10}$1/2-X,3/2-Y,1-Z; $^{11}$1-X,1-Y,1-Z}
\label{tabcrys4}
\end{table}

\begin{table}[!h]
\centering
\begin{tabular}{||c|c|c|c|c|c|c||}
\hline\textbf{Atom} & \textbf{U$_{11}$} & \textbf{U$_{22}$} & \textbf{U$_{33}$} & \textbf{U$_{23}$}&\textbf{ U$_{13}$} &\textbf{U$_{12}$}\\
\hline Rb1 & 12.3(6) & 17.9(7) & 13.1(6) & 0 & -1.6(4)& 0 \\
\hline Ti1 & 5.5(9) & 10.9(10) & 10.5(9) & 0 & -4.3(6) & 0 \\
\hline O1 & 6(3) & 16(4) & 19(3) & 0 & -6(3) & 0\\
\hline O2 &13(3) & 15(3) & 16(3) & 0 & -4(3) & 0\\
\hline O3 & 4(4) & 19(5) & 22(5) & 0 & 0(4) & 0  \\
\hline
\end{tabular}
\caption{\textbf{Anisotropic Displacement Parameters (\AA $^2 \times$10$^3$) for Rb$_2$Ti$_2$O$_5$ at 150K. The anisotropic displacement factor exponent takes the form: -2$\pi^2(h^2a^{*2}U_{11}+2hka^*b^*U_{12}+)$.}}
\label{tabcrys3}
\end{table}

\begin{table}[!h]
\centering
\begin{tabular}{||c|c|c|c|c|c|c||}
\hline \textbf{Atom} & \textbf{Atom} & \textbf{Length(\AA)} &  & \textbf{Atom} & \textbf{Atom} & \textbf{Length(\AA)} \\
\hline Rb1 & Rb1$^1$ & 3.3460(16) &  & Ti1 & O3 & 1.8910(15)\\
\hline Rb1 & Ti1$^2$ & 3.5845(17) &  & Ti1 & O3$^{8}$ & 1.9805(17)\\
\hline Rb1 & Ti1$^3$ & 3.5845(17) &  & Ti1 & O3  & 2.000(6)\\
\hline Rb1 & O1$^2$  & 3.1265(7) &  & Ti1 & O3$^9$ & 1.9805(17)\\
\hline Rb1 & O1$^3$ &3.1265(7)&  & Ti1 & O2 & 1.710(7)\\ 
\hline Rb1 & O3$^4$ &3.513(6) &  & O1	 & Rb1$^7$ & 3.1266(7)\\
\hline Rb1 & O3 &2.877(7) &  & O1 & Rb1$^{9}$ & 3.1266(7)\\
\hline Rb1 & O2$^2$ &3.154(3) &  & O3 & Ti1$^{9}$ & 1.9824(9)\\
\hline Rb1 & O2 & 3.171(6) &  & O1 & Rb1$^{8}$ & 3.1266(7)\\
\hline Rb1 & O2$^2$ &3.148(5)&  & O1 & Rb1$^{10}$ & 3.1266(7)\\
\hline Rb1 & O2$^5$ &2.878(5) &  & O1 & Ti1$^{11}$ & 1.8910(15)\\
\hline Rb1 & O2$^6$ &2.878(5) &  & O3 & Rb1$^{4}$ & 3.513(6)\\
\hline Rb1 & O2$^3$ & 3.148(5) &  & O3 & Ti$^{9}$& 1.9805(17) \\
\hline Ti1 & Rb1$^7$ & 3.5845(17) &  & O3	 & Ti1$^{8}$& 1.9805(17) \\
\hline Ti1 & Rb1$^8$ & 3.7221(16)&  & O2 & Rb1$^{5}$& 2.878(5) \\
\hline Ti1 & Rb1$^{9}$ & 3.7221(16) &  & O2 & Rb1$^{7}$& 3.148(5)\\
\hline Ti1 & Rb1$^{10}$ & 3.5845(17) &  & O2 & Rb1$^{10}$& 3.148(5) \\
\hline Ti1 & Ti1$^{9}$ & 3.091(2) &  & O2 & Rb1$^{6}$& 2.878(5) \\
\hline Ti1 & Ti1$^{8}$ & 3.091(2)	  &  &   &  &   \\
\hline
\end{tabular}
\caption{\textbf{Bond Lengths for  for Rb$_2$Ti$_2$O$_5$ at 150K. }\\$^1$-X,1-Y,2-Z; $^2$-1/2+X,-1/2+Y,+Z; $^3$-1/2+X,1/2+Y,+Z; $^4$-X,1-Y,1-Z; $^5$1/2-X,3/2-Y,2-Z; $^6$1/2-X,1/2-Y,2-Z; $^7$1/2+X,1/2+Y,+Z; $^8$1/2+X,-1/2+Y,+Z; $^9$1/2-X,1/2-Y,1-Z; $^{10}$1/2-X,3/2-Y,1-Z; $^{11}$1-X,1-Y,1-Z}
\label{tabcrys4}
\end{table}

\begin{figure*}[t]
\includegraphics[width=17cm, trim = 10cm 5cm 10.5cm 3.5cm, clip]{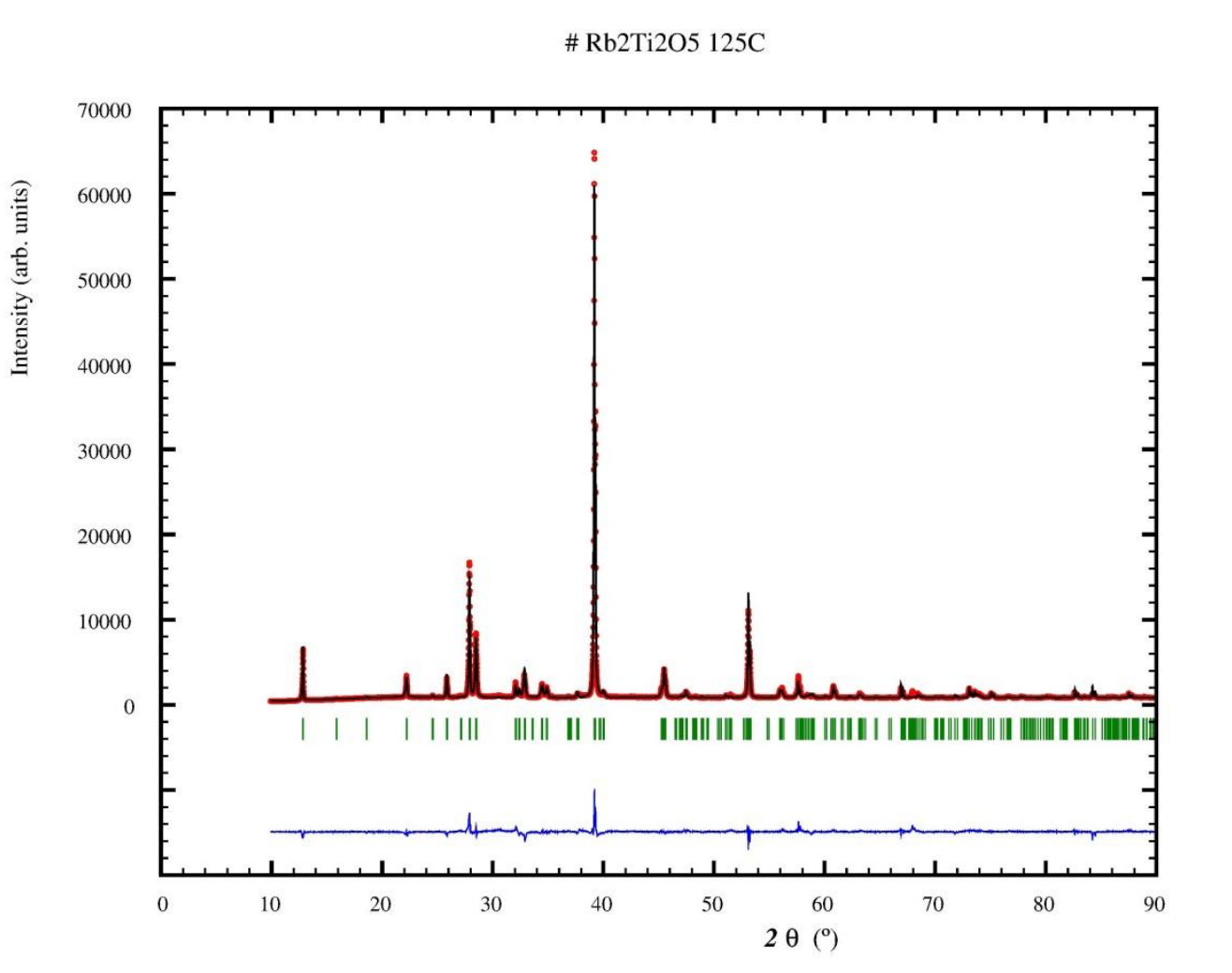}
\caption{Rietveld refinements of the Rb$_2$Ti$_2$O$_5$ X-Ray powder diffraction pattern ($\lambda_{Cu}$) at 400\,K. For all, the red circles are the experimental points, the black line are the calculated pattern, blue vertical tick marks refer to Bragg reflections and the blue line are the difference (observed-calculated) pattern. }
\label{Poudre125C}
\end{figure*}

\begin{figure*}[t]
\includegraphics[width=17cm, trim = 10cm 5cm 10.5cm 3.5cm, clip]{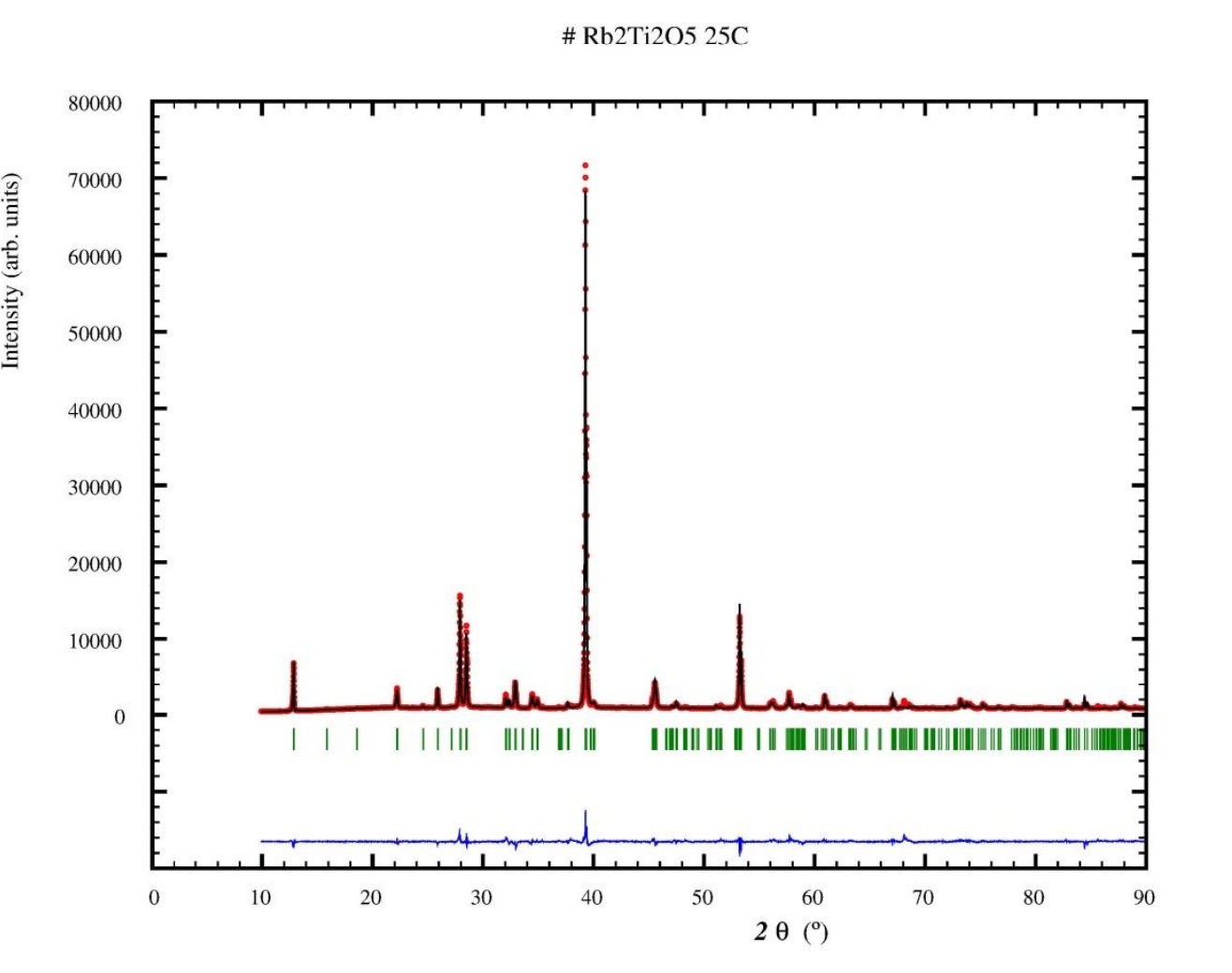}
\caption{Rietveld refinements of the Rb$_2$Ti$_2$O$_5$ X-Ray powder diffraction pattern ($\lambda_{Cu}$) at 300\,K. For all, the red circles are the experimental points, the black line are the calculated pattern, blue vertical tick marks refer to Bragg reflections and the blue line are the difference (observed-calculated) pattern. }
\label{Poudre25C}
\end{figure*}

\begin{figure*}[t]
\includegraphics[width=17cm, trim = 10cm 5cm 10.5cm 3.5cm, clip]{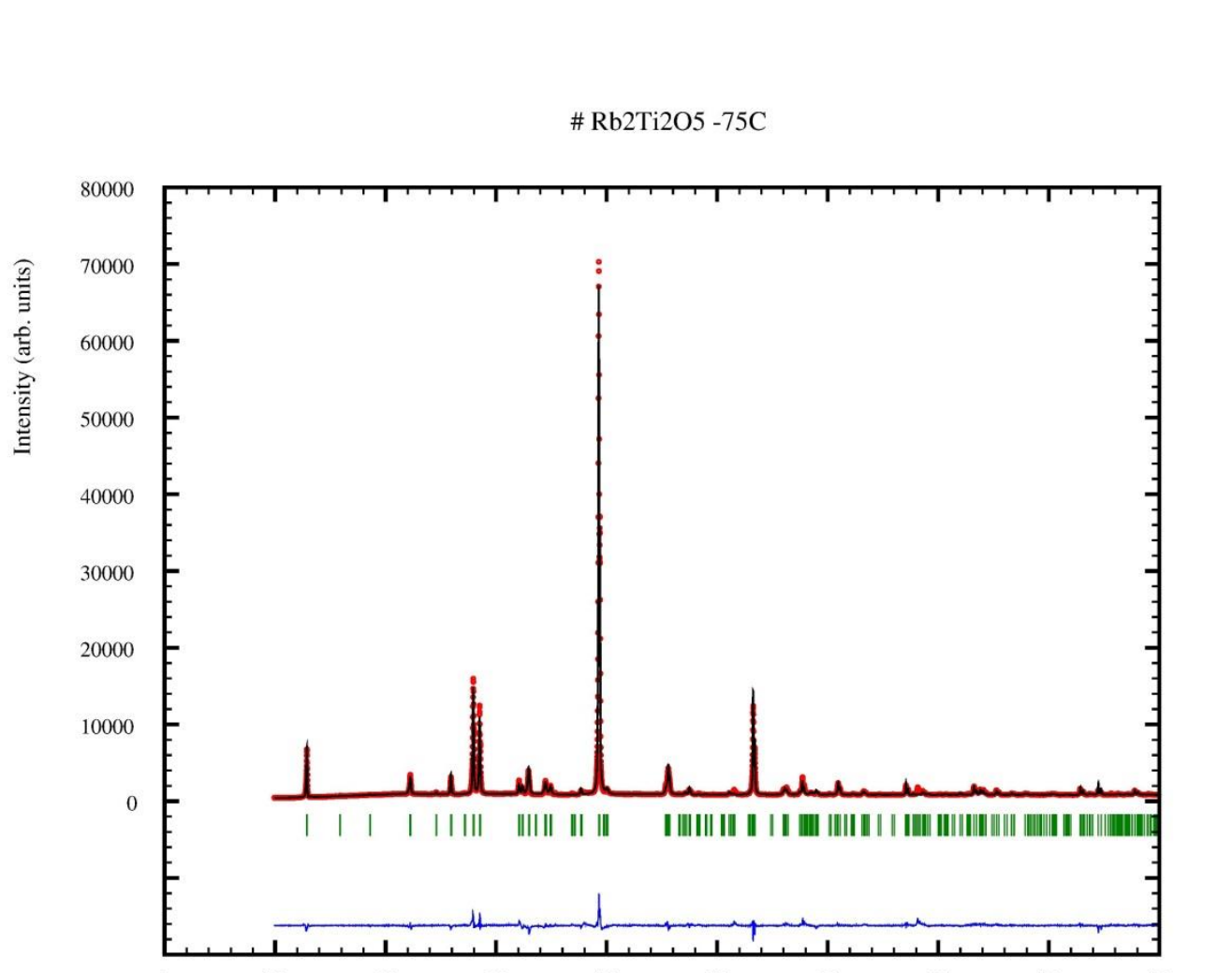}
\caption{Rietveld refinements of the Rb$_2$Ti$_2$O$_5$ X-Ray powder diffraction pattern ($\lambda_{Cu}$) at 200\,K. For all, the red circles are the experimental points, the black line are the calculated pattern, blue vertical tick marks refer to Bragg reflections and the blue line are the difference (observed-calculated) pattern. }
\label{Poudre-75C}
\end{figure*}

\begin{figure*}[t]
\includegraphics[width=17cm, trim = 10cm 5cm 10.5cm 3.5cm, clip]{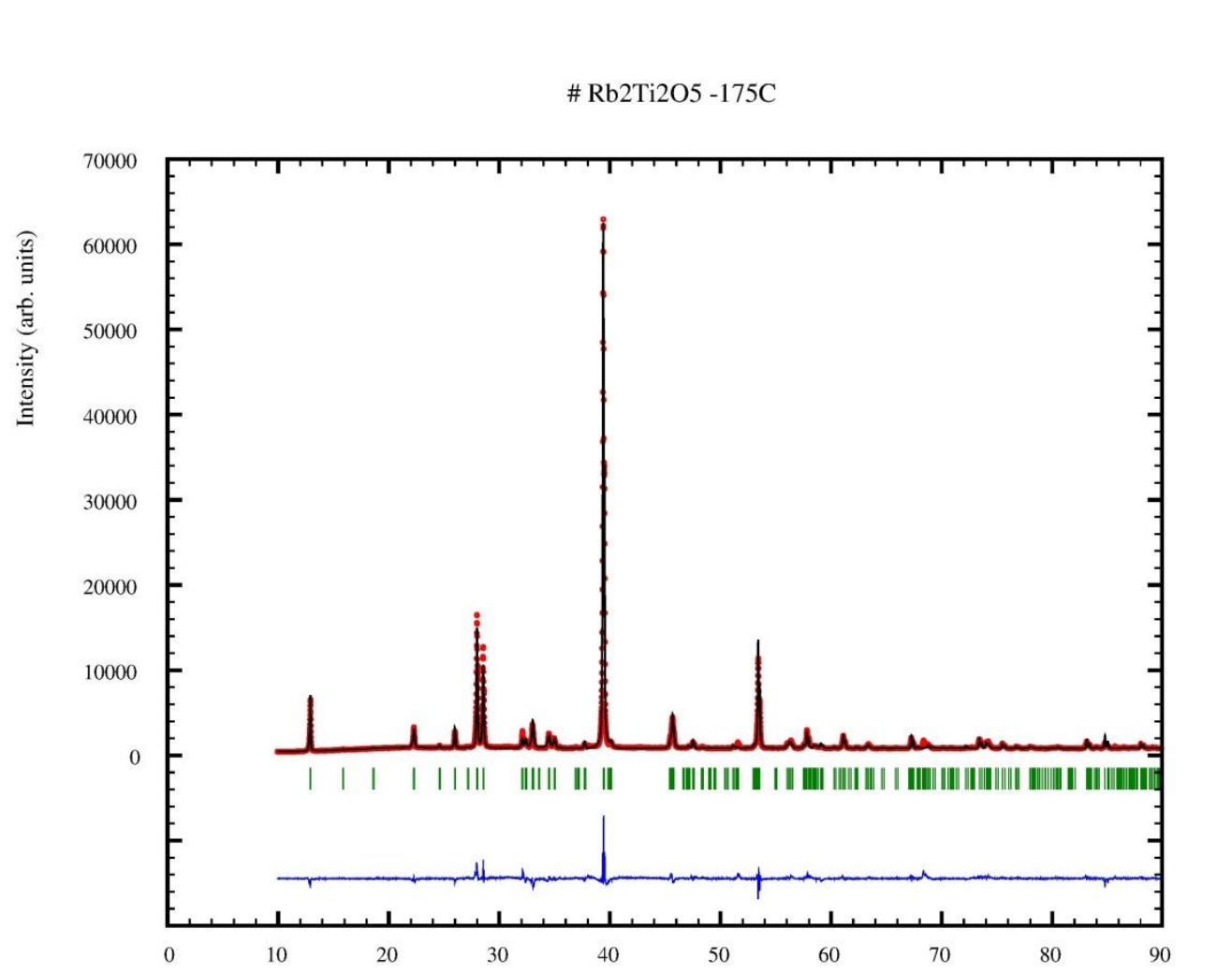}
\caption{Rietveld refinements of the Rb$_2$Ti$_2$O$_5$ X-Ray powder diffraction pattern ($\lambda_{Cu}$) at 100\,K. For all, the red circles are the experimental points, the black line are the calculated pattern, blue vertical tick marks refer to Bragg reflections and the blue line are the difference (observed-calculated) pattern.}
\label{Poudre-125C}
\end{figure*}

\section{Appendix B : representations of the DFT phonon modes calculated with GGA functional}

\begin{table}[!h]
\centering
\begin{tabular}{||c|c|c||c|c|c||}
\hline \textbf{N$\degree$} & \textbf{$\nu$ (cm$^{-1}$)} & \textbf{Raman Intensity (au)} & \textbf{N$\degree$} & \textbf{$\nu$ (cm$^{-1}$)} & \textbf{Raman Intensity (au)}\\
\hline 1 & 0  & low  & 28  & 240 & high\\
\hline 2 &  0& low & 29  &252 & medium\\
\hline 3 & 0  &  low & 30  &254 & high\\
\hline 4 & 0  &  low & 31  & 256& low\\
\hline 5 &  0 & low & 32  &282 & medium\\
\hline 6 &  0 &  low & 33  & 292& low\\
\hline 7 &  9 & low & 34  & 301& high\\
\hline 8 & 63   &  low & 35  &309 & low\\
\hline 9 &  67 &  low & 36  &319 & high\\
\hline 10 & 73  & low & 37  &387 & low\\
\hline 11 & 86  & low  & 38  & 400& low\\
\hline 12 & 87  & low  & 39  &413 & low\\
\hline 13 &  88 & low  & 40  &432 & low\\
\hline 14 & 89  & low  & 41  &444 & medium\\
\hline 15 & 92  & low & 42  &447 & high\\
\hline 16 & 115  & low  & 43  & 457& low\\
\hline 17 &  125 & low  & 44  & 498& low\\
\hline 18 &  134 & medium  & 45  &507 & high\\
\hline 19 & 145  & low  & 46  &594 & high\\
\hline 20 &  155 & low  & 47  &599 & medium\\
\hline 21 & 165  & medium & 48  &624 & low\\
\hline 22 &  183 &  low & 49  & 721& low\\
\hline 23 & 193  & low  & 50  &759 & low\\
\hline 24 & 219  & high  & 51  &759 & low\\
\hline 25 & 225  & medium  & 52  &846 & low\\
\hline 26 &  230 & medium  & 53  &879 & medium\\
\hline 27 &  234 & medium  & 54  & 881&high\\
\hline
\end{tabular}
\caption{List of vibrational phonon modes calculated with the Density Functional Theory using Generalized Gradient Approximation for the Rb$_2$Ti$_2$O$_5$ system. High Raman intensity range : [1500 - 15000], medium Raman intensity range : [150 - 1500], low Raman intensity range : [0 - 150]. The phonon mode representations are displayed in figure \ref{Phonon47-54}, \ref{Phonon39-46},  \ref{Phonon31-38}, \ref{Phonon23-30}, \ref{Phonon15-22}, \ref{Phonon7-14}, \ref{Phonon1-6}}
\label{DFT-PBE-PhononFrequency}
\end{table}

\newpage

\begin{figure*}[t]
\includegraphics[width=16cm, trim = 1cm 0.5cm 1cm 0cm, clip]{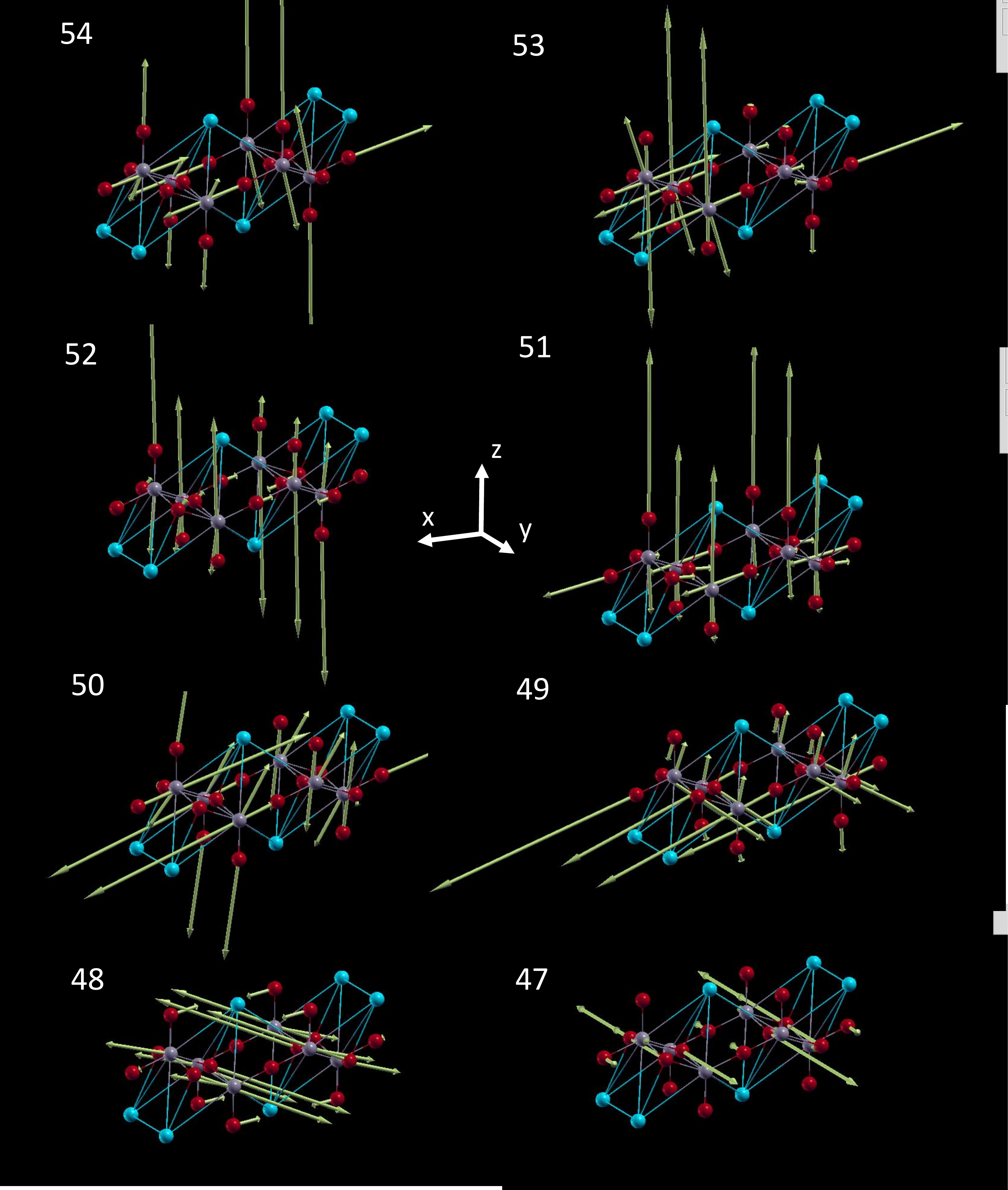}
\caption{Representations of the vibrational phonon modes N$\degree$ 54, 53, 52, 51, 50, 49, 48 and 47 calculated using Density Function Theory for Rb$_{2}$Ti$_{2}$O$_{5}$ system. The frequency and Raman intensity of the related phonon modes are displayed in the table \ref{DFT-PBE-PhononFrequency}}
\label{Phonon47-54}
\end{figure*}

\newpage

\begin{figure*}[t]
\includegraphics[width=16cm, trim = 1cm 2.5cm 1cm 0.1cm, clip]{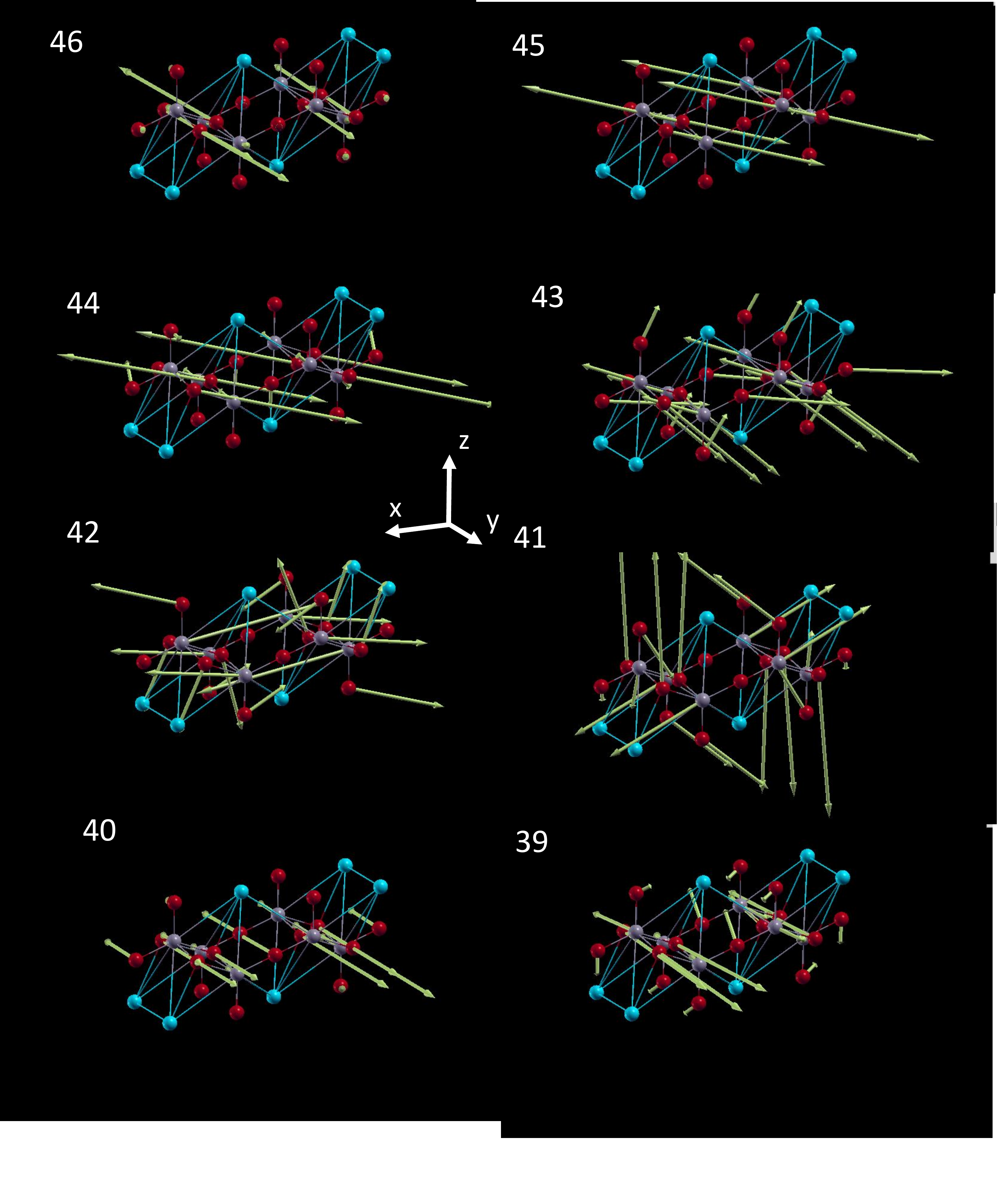}
\caption{Representations of the vibrational phonon modes N$\degree$ 46, 45, 44, 43, 42, 41, 40 and 39 calculated using Density Function Theory for Rb$_{2}$Ti$_{2}$O$_{5}$ system. The frequency and Raman intensity of the related phonon modes are displayed in the table \ref{DFT-PBE-PhononFrequency}}
\label{Phonon39-46}
\end{figure*}

\newpage

\newpage

\newpage

\newpage

\newpage

\end{document}